\documentclass[showpacs,aps,graphicx,twocolumn]{revtex4}
\usepackage{amssymb}
\usepackage{amsmath}
\usepackage{graphicx}
\usepackage{array}

\begin{document}

\title{Heralded quantum repeater for a quantum communication network based on quantum dots
embedded in optical microcavities\footnote{Published in Phys. Rev. A \textbf{93}, 012302 (2016)}}

\author{Tao Li,  Guo-Jian Yang,  and Fu-Guo Deng\footnote{Corresponding author:fgdeng@bnu.edu.cn}}
\affiliation{Department of Physics, Applied Optics Beijing Area
Major Laboratory, Beijing normal University, Beijing 100875, China}

\date{\today }

\begin{abstract}
We propose a heralded quantum repeater  protocol based on the
general interface between the circularly polarized photon and
the quantum dot embedded in a double-sided optical microcavity. Our
effective time-bin encoding on photons results  in the deterministic
faithful entanglement distribution with   one optical fiber for the
transmission of each photon in our protocol, not two or more. Our
efficient parity-check detector implemented with only one
input-output process of a single photon as a result of cavity
quantum electrodynamics makes the entanglement channel extension and
entanglement purification in quantum repeater far more efficient
than others, and it has the potential application in fault-tolerant
quantum computation as well.  Meanwhile,  the deviation from a
collective-noise channel leads to  some phase-flip errors on the
nonlocal electron spins shared by the parties and these errors can
be depressed by our simplified entanglement purification process.
Finally, we discuss the performance of  our proposal, concluding
that it is feasible with current technology.

\end{abstract}

 \maketitle

\section{introduction}

The reliable transmission of  quantum states over noisy channels is
important in  quantum communication, such as quantum teleportation
\cite{teleportation}, dense coding \cite{Densecoding,Densecoding1},
quantum key distribution \cite{QKD,QKD1,collectivenoiselixhQKD},
quantum secret sharing \cite{QSS,QSS1,QSS2}, and quantum secure
direct communication \cite{QSDC1,QSDC2,QSDC3}. However, serious
problems occur when long-distance quantum communication is
considered \cite{noise}. Due to the exponential scaling photon loss
in the transmission channel, the success probability of the direct
transmission for photons over a  $1000$-km  optical fiber is of
order $10^{-20}$. Even though the photon can arrive at the receiver,
the fidelity of its polarization state also decreases largely, due
to the random birefringence arising from thermal fluctuations,
vibrations, and imperfections of the fiber itself. To establish a
long-distance entanglement channel, a quantum repeater protocol was
originally proposed by Briegel \emph{et al.} \cite{Qrepeater0} in
1998 to reduce the photon loss rate and suppress the decoherence of
entangled photon pairs. Some interesting proposals for quantum
repeaters have been proposed in various physical systems, such as
nitrogen vacancy (NV) centers in diamond
\cite{high-frep1,high-frep2,qqentangle1}, atomic ensembles
\cite{repeaternature,repeaterPRL,revqr}, and single trapped ions
\cite{ionqr}.

Considering the long electron-spin coherence time [$\mu$s is
achieved in both a quantum dot (QD) ensemble and  a single QD], fast
manipulation, and easy scalability, QD is one of the good candidates
for local storage and processing of quantum information. Single
semiconductor QD  coupling to a microcavity has attracted much attention
\cite{Qdcoupling3,weihairuiOE2,wanghongfuPRA,
RenbaocangSR,renHEPPPRA,Qdcoupling4,Qdcoupling0,Qdqr1,Qdqr2,Qdqr3,qqentangle2}.
The giant circular birefringence originated from the spin selective
dipole coupling for such spin-cavity systems is utilized in
photon-photon or spin-photon entanglement generation
\cite{Qdcoupling3}, hyper-parallel quantum computing
\cite{RenbaocangSR}, universal quantum gates
\cite{Qdcoupling0,weihairuiOE2,wanghongfuPRA}, hyperentanglement purification and
concentration \cite{renHEPPPRA}, and complete Bell-state analyzers
\cite{Qdcoupling4}. In 2006, Waks and Vuckovic put forward a quantum
repeater scheme with the  QD-cavity system \cite{Qdqr1}. The entanglement
creation between neighboring QDs and the subsequently  entanglement
swapping  were assisted by the QD-induced transparency of the coherent
field, which is faithful when the photon number resolved detectors
were available.  In 2007, Simon \emph{et al.} \cite{Qdqr2} proposed
a scheme for entangling two remote spins based on two-photon
coincidence detection and they constituted a controlled-phase gate
between two local spins  with the dipole-dipole interaction between
trions in neighboring QDs. This gate makes the quantum entanglement
swapping possible and it leads to the realization of a quantum
repeater. In 2012, the error-free entanglement distribution was
performed with the momentum-entangled  photons and the QDs embedded
in microcavities when the momentum entanglement is stable
\cite{Qdqr3}.  Recently, Jones \emph{ et al.} \cite{qqentangle2}
proposed an  efficient scheme to entangle remotely separated  QDs
with a midpoint-entanglement source and one nondeterministic
Bell-state measurement  located at each end of the two channels,
which was used to complete the entanglement swapping between the
QD-photon entanglement \cite{Qdphoton1,Qdphoton2,Qdphoton3} and the
photon-photon entanglement, resulting in the entanglement between
QDs.

Since the seminal work about Bell inequality for position and time
by Franson \cite{time-bin1}, the time-bin degree of freedom (DOF) of
photons has attracted much attention
\cite{time-bin2,time-bin3,lixhapl,time-bin4,time-bin5,time-bin6,time-bin7}. The
two-photon time-bin entanglement source for quantum communication
was demonstrated by Brendel \emph{et al.} \cite{time-bin2} in 1999.
With the encoded time-bin qubits, Kalamidas
\cite{time-bin3} proposed a single-photon quantum  
error-rejection transmission protocol in 2005, in which a
probabilistic transmission is completed with two Pockels cells (PCs)
and the deterministic error-free transmission is performed with four
PCs. In 2007, Li \emph{et al.} \cite{lixhapl} proposed a faithful
qubit transmission scheme against collective noise without ancillary
qubits, resorting to the time-bin DOF of a single photon itself.
Recently, the distribution of time-bin entangled qubits
\cite{time-bin4} over an optical fiber at the scale of 300 km
\cite{time-bin5} was demonstrated and the two-photon interference
fringes  exhibited  a visibility of 84\%. A time-bin qubit can also
be used to perform quantum computing \cite{time-bin6} and only a
single optical path rather than multiple paths is used to complete
single-qubit operations and herald controlled-phase gates. An
ultrafast measurement technique for time-bin qubits
\cite{time-bin7} was implemented, which makes time-bin qubits  more
useful \cite{time-bin8}.

In this paper,  we show that a heralded  quantum repeater based on
the QD-microcavity systems can be constructed   with the help of the
effective time-bin encoder and the general interface between the
circularly polarized photon and the QD embedded in a double-sided
optical microcavity.  By using the giant circular birefringence
effect for the singly charged QD inside a microcavity and the
two-photon coincident measurement, the time-bin entanglement can be
converted deterministically into that of the remotely located
QD-electron-spin  system in a heralded way. The entanglement
distribution can in principle be performed with a unity efficiency when none of
the photons are lost during the transmission process. It is more
efficient than others \cite{Qdqr2,qqentangle2} if the multimode
process is involved \cite{QRmulti-spatial,qqentangle2}. Our
efficient parity-check detector (PCD)  implemented with only one
input-output process of a single photon as a result of cavity
quantum electrodynamics makes the entanglement channel extension and
entanglement purification in our quantum repeater far more efficient
than others. The deviation from collective noise channel leads to
some phase-flip errors that can be suppressed by our simpler
entanglement purification process. These features make our heralded
quantum repeater protocol more useful in the quantum communication
network in the future.

This paper is organized as follows: We give a general interface
between a circularly polarized light and a QD-cavity system in Sec.
\ref{sec2A}. Subsequently,  we present the faithful entanglement
distribution for two neighboring nodes in Sec. \ref{sec2B}, and then, we give an efficient way to complete the entanglement
extension with a PCD  in
Sec. \ref{sec2D}.  In Sec.  \ref{sec3}, we propose an efficient
entanglement purification protocol to depress the influence of
asymmetric noise from optical-fiber channels on different time bins.
In Sec. \ref{sec4}, we discuss the influence from the practical
imperfect circular birefringence on the created entanglement. A
discussion and a summary are given in Sec. \ref{sec5}. In addition, $N$-user entanglement
distribution for a multiuser
quantum repeater network is discussed in  Appendix \ref{AppendixA}.

\begin{figure}[!h]
\begin{center}
\includegraphics[width=6.4 cm,angle=0]{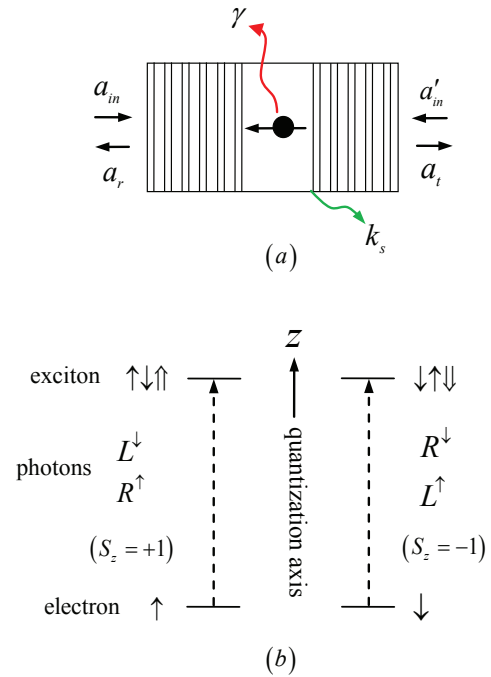}
\caption{The spin-dependent transitions for a negatively charged
exciton $X^-$. (a) A singly charged QD inside a double-sided optical
microcavity. (b) The spin selection rules for optical transition of
a  negatively charged exciton.  The symbols $\uparrow$ and
$\downarrow$ represent the excess electron-spin  projections
$|+\frac{1}{2}\rangle$ and $|-\frac{1}{2}\rangle$ along the
quantization axis ($z$-direction), respectively. The symbols
$\Uparrow$ and $\Downarrow$ represent the spin projections of the
hole $|+\frac{3}{2}\rangle$ and $|-\frac{3}{2}\rangle$,
respectively. $R^\uparrow$ ($L^\downarrow$) denotes a
right-circularly (a left-circularly) polarized photon propagating
along (against) the quantization axis.} \label{fig1}
\end{center}
\end{figure}

\section{faithful entanglement distribution and extension for heralded quantum repeater}

\subsection{ The interface between a circularly polarized light and a
QD-cavity system } \label{sec2A}

Let us consider a singly charged QD (e.g., for a self-assembled
InAs/GaAs quantum dot) embedded inside a  resonant double-sided
micropillar cavity \cite{Qdcoupling3}. Both the top and bottom
mirrors of the cavity are partially reflective, shown in
Fig. \ref{fig1}(a). The optical properties of a singly charged QD
embedded inside a micropillar cavity are dominated by the optical
transitions of the negatively charged trion ($X^-$) that consists of
two electrons bounded to one hole \cite{Qdchar}, shown in
Fig.\ref{fig1}(b). When the quantization axis for angular momentum
is the $z$ axis for the QD geometry, the single electron
states have the spin $J_z =\pm \frac{1}{2}$ (labeled as
$|\uparrow\rangle$ and $|\downarrow\rangle $), and the hole $J_z
=\pm \frac{3}{2}$ (labeled as $|\Uparrow\rangle$ and
$|\Downarrow\rangle$). Photon polarization ($L$ and $R$ represent
the left and the right circularly polarized states of photons,
respectively) is defined with respect to the direction of the
propagation, and this causes the polarization to change upon
reflection. In a trion state, due to the Pauli's exclusion
principle, the two electrons form a singlet state with the total
spin zero, which decouples the interaction between the electron spin
and the hole spin \cite{Qdchar,Qdchartri}. In other words, the
circularly polarized photon directed into the spin-cavity system can
either be coupled with the electron spin and feels a hot cavity when
the dipole selection rule is fulfilled, or be decoupled and feels a
cold cavity in the other case. The significant difference in the
reflection and the transmission coefficients manifested between
these two cases is spin dependent, and it can be exploited to perform the quantum information processing \cite{Qdcoupling3,weihairuiOE2,wanghongfuPRA,
RenbaocangSR,renHEPPPRA,Qdcoupling4,Qdcoupling0,Qdqr1,Qdqr2}.

The reflection and the transmission coefficients  of this
spin-cavity system can be obtained by solving the Heisenberg
equations of motion for the cavity field operator $\hat{a}$  and the
trion dipole operator $\hat{\sigma}_-$  along with  the input-output
relations
\cite{Qdqr1,doubleEq,singleEq}.
\begin{eqnarray}   
\begin{split}
\frac{d \hat{a}}{d t}\;=\;\;&-\left( i\omega_c+\kappa+\frac{\kappa_s}{2}\right)\hat{a}-ig\hat{\sigma}_- \\
&-\sqrt{\kappa_s}\,\hat{s}_{in}-\sqrt{\kappa}\,\hat{a}_{in}-\sqrt{\kappa}\,\hat{a}_{in}',\\
\frac{d \hat{\sigma}_-}{d t}\;=\;\;&-\left( i\omega_{X^-}+\frac{\gamma}{2} \right)\hat{\sigma}_-
+ig\hat{\sigma}_z\hat{a}+\sqrt{\gamma}\,\hat{\sigma}_z\hat{N},\;\;\;\;\;\;\\
\hat{a}_r\;=\;\;&\hat{a}_{in}+\sqrt{\kappa}\,\hat{a},\\
\hat{a}_t\;=\;\;&\hat{a}_{in}'+\sqrt{\kappa}\,\hat{a}.
\end{split}
\end{eqnarray}
Here $\hat{a}_{in}$ and $\hat{a}_{in}'$ are the two input field
operators. While $\hat{s}_{in}$ is an operator for an input field
originating from potential leaky modes due to sideband leakage and
absorption, and it associates with the corresponding output mode
$\hat{s}_{out}$ by $\hat{s}_{out}
=\hat{s}_{in}+\sqrt{\kappa_s}\hat{a}$ . $\hat{a}_r$ and $\hat{a}_t$
are the two output field operators, shown in Fig.\ref{fig1}(a).
$\hat{N}$ is the corresponding vacuum noise operator which helps to
preserve the desired commutation relations for the QD-dipole
operators. $\omega$, $\omega_c$, and $\omega_{X^-}$ are the
frequencies of the input photon, cavity mode, and $X^-$ transition,
respectively.  $\kappa$ and $\kappa_s$ are the cavity-field decay
rate  and the side-leakage rate, respectively. $g$ is the coupling
strength between $X^-$ and the cavity mode. $\gamma/2$ is the dipole
decay rate.  In the limit of weak incoming field, the charged QD is
predominantly in the ground state in the whole process, that is,
$<\hat{\sigma}_z> \approx - 1$. The spin-cavity system behaves like
a beam splitter whose reflection  and transmission  coefficients
$R(\omega)$ and $T(\omega)$  along with the leakage and noise
coefficients $S(\omega)$ and $N(\omega)$  are detailed,
respectively, by:
\begin{eqnarray}   
\begin{split}
R(\omega)\;\;=\;\;&\frac{i(\omega_c-\omega)+\frac{\kappa_s}{2}+\frac{g^{2}}{i(\omega_{X^-}-\omega)
+\frac{\gamma}{2}}}{i(\omega_c-\omega)+\kappa+\frac{\kappa_s}{2}+\frac{g^{2}}{i(\omega_{X^-}-\omega)+\frac{\gamma}{2}}},\\
T(\omega)\;\;=\;\;&\frac{-\kappa}{i(\omega_c-\omega)+\kappa+\frac{\kappa_s}{2}+\frac{g^{2}}{i(\omega_{X^-}-\omega)+\frac{\gamma}{2}}},\;\;\;\;\;\;
 \label{hotrt11}
\end{split}
\end{eqnarray}
and
\begin{eqnarray}
\begin{split}
S(\omega)\;\;=\;\;&\frac{-\sqrt{\kappa_s\kappa}}{i(\omega_c-\omega)+\kappa+\frac{\kappa_s}{2}+\frac{g^{2}}{i(\omega_{X^-}-\omega)+\frac{\gamma}{2}}},\\
N(\omega)\;\;=\;\;&\frac{\frac{ig\sqrt{\gamma\kappa}}{i(\omega_{X^-}-\omega)+\frac{\gamma}{2}}}{i(\omega_c-\omega)
+\kappa+\frac{\kappa_s}{2}+\frac{g^{2}}{i(\omega_{X^-}-\omega)+\frac{\gamma}{2}}}.\;\;\;\;\;\;
 \label{hotrt0}
\end{split}
\end{eqnarray}
Here,  one has
$P(\omega)=|R(\omega)|^2+|T(\omega)|^2+|S(\omega)|^2+|N(\omega)|^2=1$ meaning
that when the noise or environment  is considered, the energy of the
whole system is conserved during the input-output process
described above, and it can be reduced to the simplified
input-output models in \cite{doubleEq}  or   \cite{Qdqr1}  by omitting
the leaky modes $\hat{s}_{in}$  or  vacuum noise operator $\hat{N}$
that helps to preserve the desired commutation relations for the
QD-dipole operators, respectively.

\begin{figure}[t]
\includegraphics[width=8.4 cm]{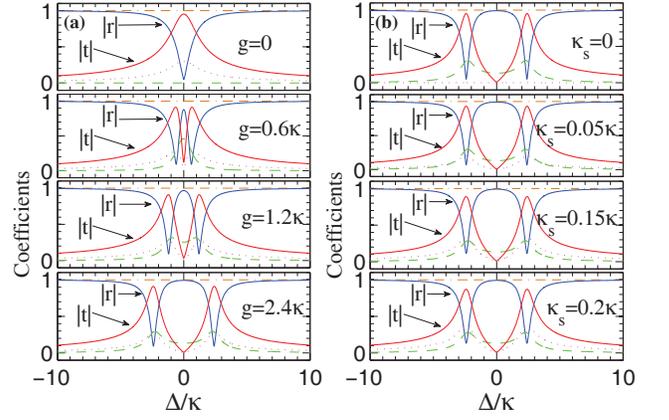}
\caption{(Color online)  The coefficients $|R(\omega)|$,
$|T(\omega)|$,  $|S(\omega)|$, $|N(\omega)|$, and $P(\omega)$ vs
detuning $\Delta/\kappa$  (a) for different coupling strengths ($g=0$,
$0.6\kappa$, $1.2\kappa$, and $2.4\kappa$) with
$\kappa_s/\kappa=0.1$ and  (b) for different  leakage rates
($\kappa_s=0$, $0.05\kappa$, $0.15\kappa$, and $0.2\kappa$) with
$g/\kappa=2.4$. $|R(\omega)|$ : solid blue line;  $|T(\omega)|$:
solid red line;  $|S(\omega)|$: dot magenta  line; $|N(\omega)|$:
dashed green line;  $P(\omega)$: dash-dash-dot brown line.
Here $\Delta=\omega_c-\omega$. $\gamma/\kappa=0.1$ is taken by  considering  the typical QD micropillar parameters.}\label{fig2}
\end{figure}

We are interested in the reflection and transmission of a single
input photon that leads to the click of a single-photon detector,
while the environment excitation inhibits the click. Therefore, we
can project the output photon into the subspace spanned by the
reflection and transmission modes. The state vector evolves on
reflection and transmission as
\begin{eqnarray}
\hat{a}_{in}^{\dagger}|S\rangle\;\rightarrow\;R\,\hat{a}_r^{\dagger}|S\rangle+T\,\hat{a}_t^{\dagger}|S\rangle.
 \label{hotrt1}
\end{eqnarray}
For  simplification, we take the case that the trion dipole is tuned
into the cavity mode ($\omega_c=\omega_{X^-}$). When the input
photon couples to the QD embedded in the microcavity, the
coefficients $R$ and $T$ are reduced to $r(\Delta)$ and $t(\Delta)$, respectively. Here
\begin{eqnarray}   
\begin{split}
r(\Delta)\;\;=\;\;&\frac{i\Delta+\frac{\kappa_s}{2}+\frac{g^{2}}{i\Delta
+\frac{\gamma}{2}}}{i\Delta+\kappa+\frac{\kappa_s}{2}
+\frac{g^{2}}{i\Delta+\frac{\gamma}{2}}}, \\
t(\Delta)\;\;=\;\;&\frac{-\kappa}{i\Delta+\kappa+\frac{\kappa_s}{2}
+\frac{g^{2}}{i\Delta+\frac{\gamma}{2}}},
 \label{hotrt}
\end{split}
\end{eqnarray}
where $\Delta=\omega_c-\omega$. When the input probe field is
uncoupled to the dipole transition (i.e., $g=0$), the specific
reflection and transmission coefficients can be simplified as
\begin{eqnarray}   
\begin{split}
r_0(\Delta)\;\;=\;\;&\frac{i\Delta+\frac{\kappa_s}{2}}{i\Delta+\kappa
+\frac{\kappa_s}{2}}, \\
t_0(\Delta)\;\;=\;\;&\frac{-\kappa}{i\Delta+\kappa+\frac{\kappa_s}{2}}.
 \label{coldrt}
\end{split}
\end{eqnarray}

For the condition  $\Delta=0$, ${2g^2}/{\kappa\gamma}\gg1$, and
${\kappa_s}/{2\kappa}\ll 1$, both the reflection coefficient
$|r(\omega)|$ and the transmission coefficient $|t_0(\omega)|$ can
approach $1$. Meanwhile, the noise terms
$|S(\Delta)|\simeq\sqrt{\kappa_s/\kappa}$ and
$|N(\Delta)|\simeq{}$$\sqrt{\kappa\gamma}/g$ can be neglected.
However, the total probabilities for all channels that the input photon is scattered into by the QD-cavity system
$P(\Delta)\equiv1$ for any condition, shown in Fig.
\ref{fig2}. To be exact, when we concern only the transmission and
reflection modes, and the circularly polarized photon directed into
the spin-cavity system is in the state $S_z=+1$ (i.e.,
$|L^{\downarrow}\rangle$ or $|R^{\uparrow}\rangle$), the excess
electron in the state $|\uparrow\rangle$ will interact with the
input photon, provide a hot cavity situation, and eventually make
the photon be reflected. Upon reflection, both the polarization and
the propagation direction of the photon will be flipped. However, if
the input photon is in the state $|R^{\downarrow}\rangle$ or
$|L^{\uparrow}\rangle$ ($S_z=-1$), it will be transmitted through
the cavity and acquires an extra $\pi$ phase, leaving the electron
spin state unaffected. The whole process can be summarized into the
following transformations \cite{Qdcoupling0}:
\begin{eqnarray}
\begin{split}
|R^{\uparrow},\uparrow\rangle \;\rightarrow\;&
|L^{\downarrow},\uparrow\rangle,\;\;\;\;
|R^{\downarrow},\uparrow\rangle \;\rightarrow\; -|R^{\downarrow},\uparrow\rangle, \\
|L^{\downarrow},\uparrow\rangle \;\rightarrow\;&
|R^{\uparrow},\uparrow\rangle,\;\;\;\; |L^{\uparrow},\uparrow\rangle
\;\rightarrow\; -|L^{\uparrow},\uparrow\rangle.
\label{transup}
\end{split}
\end{eqnarray}
When the excess electron is in the state $|\downarrow\rangle$, the
evolution can be described as \cite{Qdcoupling0}:
\begin{eqnarray}
\begin{split}
|R^{\uparrow},\downarrow\rangle \; \rightarrow\;&
-|R^{\uparrow},\downarrow\rangle,\;\;\;\;\;\;\;\;
|R^{\downarrow},\downarrow\rangle \;\rightarrow\;
|L^{\uparrow},\downarrow\rangle, \\
|L^{\downarrow},\downarrow\rangle \; \rightarrow\;&
-|L^{\downarrow},\downarrow\rangle,\;\;\;\;\;\;\;\;\,
|L^{\uparrow},\downarrow\rangle \;\rightarrow\;
|R^{\downarrow},\downarrow\rangle.\label{transdown}
\end{split}
\end{eqnarray}
Combining  the rules above, we can  accomplish the entanglement
transfer from the nonlocal photon systems into the nonlocal spin
systems and construct an efficient PCD which  is essential
for the entanglement purification of  the spin systems and the
entanglement extension in our quantum repeater protocol.

\begin{figure}[!ht]
\includegraphics[width=8.0 cm]{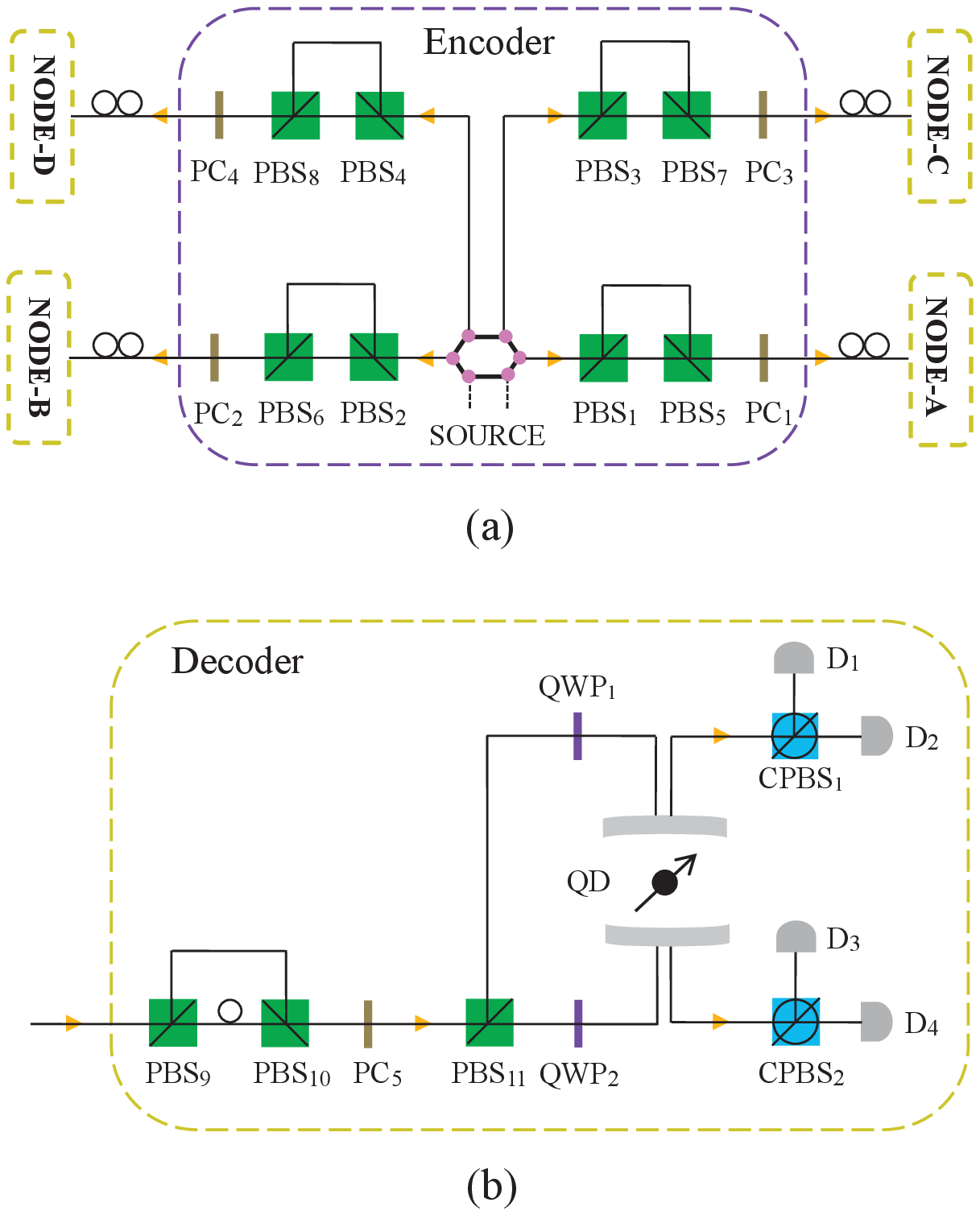}
\caption{(Color online)  Schematic architecture of the faithful
entanglement distribution procedure in our quantum repeater
protocol. (a) Deterministic and faithful entanglement distribution
between the  nodes  A and B (A, B, $\cdots$, C and D) with the help
of time-bin encoders. (b) The decoder for each quantum node. Here,
the hexagon denotes the two-photon Bell source ($N$-photon GHZ
source) and the orange rounded rectangles denoted with NODE-A and
NODE-B represent the quantum nodes owned by the users Alice and Bob,
respectively. PBS$\,_i$ ($i=1,2,\cdots$) is the polarizing beam
splitter that transmits the $|H\rangle$ polarized photon and
reflects the $|V\rangle$ polarized photon, respectively.  QWP$_j$
($j=1,2$) represents a quarter-wave plate which is used to
accomplish the transformations $|H\rangle\leftrightarrow|R\rangle$
and $|V\rangle\leftrightarrow|L\rangle$.  CPBS$_j$ represents the
circularly polarizing beam splitter that transmits the $|R\rangle$
polarized photon and reflects the $|L\rangle$ polarized photon,
respectively. PC$\,_i$ is a Pockels cell.}\label{fig3}
\end{figure}

\subsection{Faithful entanglement distribution for the quantum repeater network with time-bin encoders}
\label{sec2B}

To show how the spin-cavity system works for our deterministic
entanglement distribution and our simplified entanglement purification protocol for quantum
systems in a  mixed entangled state explicitly, we
first take the two-photon Bell state as an example, and then
generalize it to the case with an $N$-photon Greenberger-Horne-Zeilinger (GHZ) state in Appendix \ref{AppendixA}.

Suppose that there is a two-photon entangled source in the middle
point of the two memory nodes belonging to the users, say Alice and
Bob, shown in Fig. \ref{fig3} (a). The photons $ab$ produced by the
source are entangled in a Bell state in the polarization degree of
freedom (DOF), i.e.,
\begin{eqnarray}
|\Phi^+_2\rangle_s=
\frac{1}{\sqrt{2}}\left(|H\rangle_a|H\rangle_b+|V\rangle_a|V\rangle_b\right),
\label{three-initial}
\end{eqnarray}
where  $|H\rangle$ and $|V\rangle$ represent the horizontal and the
vertical polarization modes of  photons, respectively. The
subscripts  $a$ and $b$ represent the photons transmitted to Alice
and Bob,  respectively.  Before entering the noise channels, the
photonic polarization entanglement is converted into the time-bin
entanglement of the two photons $ab$ by passing the two photons
through the encoders placed in their respective paths. Each
encoder is made of two  PBSs  and a fast Pockels cell (PC). A PBS
transmits the $|H\rangle$ polarized photon and reflects  the
$|V\rangle$ polarized one.  A relative time
delay  $\Delta t$  of the nanoseconds scale can be obtained for the
$|V\rangle$ component of the photon when the users appropriately
preset the difference between  the long optical path length  $l$  of
the $|V\rangle$ polarization photon and the short optical path
length $s$  of  the $|H\rangle$ polarization photon (i.e., the time
interval of the unbalanced Mach-Zehnder interferometers).  The
parties turn on PC only when  the $l$-path component appears and it
is used to implement  the bit-flip operation
$|V\rangle\leftrightarrow|H\rangle$.  After the encoders, the state
of the system composed of  photons $ab$ is changed into the time-bin
entanglement with the polarization states all being $|H\rangle$,
\begin{eqnarray}
|\Phi^+_2\rangle_{t_0}\!=\!
\frac{1}{\sqrt{2}}|H\rangle_a|H\rangle_b
(|s\rangle|s\rangle
\!+\!|l\rangle|l\rangle)_{ab}. \label{threet-t}
\end{eqnarray}
Here $|s\rangle$ and $|l\rangle$ denote the early and late time bins
with which the photon passes through the optical short ($s$) and
long ($l$) paths, respectively.

Since all the photons in both $|s\rangle$ and $|l\rangle$ time
bins launched into the noisy channels are in the $|H\rangle$
polarization, the influences of the collective noise on the photons
in different time bins can be taken to be the same one
\cite{time-bin3,lixhapl,collectivenoise,Yamamoto05,YamamotoExp2}.
In other words, the noise of each optical-fiber channel is stable in
the nanosecond scale that is just the time separation between the
$|s\rangle$ and $|l\rangle$ time bins, and it can be expressed by a
unitary transformation $U_i$,
\begin{eqnarray}
U_i|H\rangle_i=\delta_i|H\rangle_i+\eta_i|V\rangle_i.
\label{U-i}
\end{eqnarray}
Here $|\delta_i|^2+|\eta_i|^2=1$.   $i$  ($=a, b$)  describes  the
noise  on  the  photon  $i$.  The  state  of  the photonic system
$ab$  arriving at the two nodes, i.e., Alice and Bob, can be written
as
\begin{eqnarray}
|\Phi^+_2\rangle_{t_1}\!\!&=&\!\!
\frac{1}{\sqrt{2}}(\delta_a|H\rangle_a\!+\!\eta_a|V\rangle_a)(\delta_b|H\rangle_b\!+\!\eta_b|V\rangle_b)\nonumber\\
&&\!\!\!\otimes(|s\rangle_a|s\rangle_b\!+\!|l\rangle_a|l\rangle_b),
\label{threet-noise}
\end{eqnarray}
which is still a two-photon time-bin entanglement but the
polarization state of the  photonic system is ambiguous since the unitary
transformation $U_i$ ($i=a,b$) on the photon  $i$ is arbitrary and
unknown for the parties in a quantum repeater.

The nodes NODE-A and NODE-B represent the two parties, Alice and
Bob, respectively. They have the same device setting  for their
decoders, shown in Fig. \ref{fig3} (b). After the photons pass
through the  unbalanced Mach-Zehnder interferometer composed of
PBS$_9$ and PBS$_{10}$, a relative delay of $\Delta{}t$ on the
$|H\rangle$ component is completed  and  the  state  of  the
photonic  system evolves into
\begin{eqnarray}
|\Phi^+_2\rangle_{t_2}\!\!&=&\!\!
\frac{1}{\sqrt{2}}[ \delta_a\delta_b|H\rangle_a|H\rangle_b(|sl\rangle_a|sl\rangle_b\!+\!|ll\rangle_a|ll\rangle_b)\nonumber\\
&&\!\! +\delta_a\eta_b|H\rangle_a|V\rangle_b(|sl\rangle_a|ss\rangle_b\!+\!|ll\rangle_a|ls\rangle_b)\nonumber\\
&& \!\!+\eta_a\delta_b|V\rangle_a|H\rangle_b(|ss\rangle_a|sl\rangle_b\!+\!|ls\rangle_a|ll\rangle_b)\nonumber\\
&&
\!\!+\eta_a\eta_b|V\rangle_a|V\rangle_b(|ss\rangle_a|ss\rangle_b\!+\!|ls\rangle_a|ls\rangle_b)],\;\;\;\;
\end{eqnarray}
and it is a partially polarized entangled state. The time-bin information
heralds the polarization state of the photonic system, which can be
utilized to correct the polarization error in the decoder with
proper bit-flip operations. Here $|ij\rangle=|i\rangle|j\rangle$ and
the notations $|i=s,l\rangle$ ($|j=s,l\rangle$) denote the
corresponding time bins created in the encoder (decoder).   PC$_i$
at each node is supposed to be active only when the components of
$|ls\rangle$ or $|s\,l\rangle$ time bins appear and implements the
bit flip $|H\rangle\leftrightarrow|V\rangle$ for the components
$|ls\rangle$ and $|s\,l\rangle$.  PBS$_{11}$ transmits the $|H\rangle$
components and reflects the $|V\rangle$ components, respectively.
Another relative time delay $\Delta{}t$   is exerted on the $|V\rangle$ components by setting a
longer optical path for the $|V\rangle$ components. The state
$|\Phi^+_2\rangle_{t_2}$ evolves to
\begin{eqnarray}
|\Phi^+_2\rangle_{t_3}&\!\!=\!\!&
\frac{1}{\sqrt{2}}(|H^{\uparrow}\rangle_a|H^{\uparrow}\rangle_b
\!+\!|V^{\downarrow}\rangle_a|V^{\downarrow}\rangle_b)\nonumber\\
&&\otimes[(\delta_a|s'\rangle_a\!+\!\eta_a|l'\rangle_a)
(\delta_b|s'\rangle_b\!+\!\eta_b|l'\rangle_b)].
\label{threet-input}
\end{eqnarray}
Here the superscripts ${\downarrow}$ and ${\uparrow}$ represent the
different outputs of  PBS$_{11}$, and ${\uparrow}$ is coincident with
the relative orientation of the quantization axis of the QD-confined
spin. $|s'\rangle$ ($=|ss\,l\rangle$, $|s\,ls\rangle$, or
$|lss\rangle$) denotes the time-bin component with only one delay
interval. $|l'\rangle$ ($=|s\,ll\rangle$, $|ls\,l\rangle$, or
$|lls\rangle$) denotes the time-bin component with two delay
intervals. One can easily find that no matter what time bins the
photons occupy, they are maximally entangled in the polarization
DOF.

Now, we only discuss the case
$|\Phi^+_2\rangle_t=\frac{1}{\sqrt{2}}(|H^{\uparrow}\rangle_a|H^{\uparrow}\rangle_b
+|V^{\downarrow}\rangle_a|V^{\downarrow}\rangle_b)
\otimes|s'\rangle_a|s'\rangle_b$ for the deterministic entanglement creation of
the nonlocal two-electron-spin system shared by Alice and Bob, and
the other cases can be discussed in a similar way.


To entangle the two QD-confined electron spins $e_a$ and $e_b$ owned
by Alice and Bob, respectively, they first introduce a $\pi$ phase
shift on the $|V\rangle$ component of the photon $b$ sent to Bob,
and then, they  map  the linearly polarized photon into the
circularly polarized one $|H\rangle\leftrightarrow|R\rangle$ and
$|V\rangle\leftrightarrow|L\rangle$ with the quarter-wave plates
(QWPs) near the two input ports of the cavity.  The state of the
entangled photons evolves into
$|\Phi^-_{2}\rangle_c=(|R^{\uparrow},R^{\uparrow}\rangle-|L^{\downarrow},L^{\downarrow})/\sqrt{2}$.
Before the arriving of the photons, each of the QD-confined-electron
spins $e_i$ ($i=a,b$) is initialized to be a superposition state
$|\Phi\rangle_{e_i}=\frac{1}{\sqrt2}(|\uparrow\rangle+|\downarrow\rangle)$.
With the giant optical circular birefringence induced by a single
electron QD embedded in a micropillar cavity  [see Eqs.
(\ref{transup}) and (\ref{transdown}) for detail],  the  state of
the hybrid photon-spin system after the reflection or transmission
of the photons $ab$, can be divided into two subspaces: (1) Both
photons $a$ and $b$  suffer a bit-flip or a unity operation when the
spins  $e_1$ and $e_2$ are in the same state
$|\uparrow\rangle\otimes|\uparrow\rangle$ or
$|\downarrow\rangle\otimes|\downarrow\rangle$; (2) only one of
photons $a$ and $b$ suffers a bit-flip when the spins  $e_1$ and
$e_2$ are in different states
$|\uparrow\rangle\otimes|\downarrow\rangle$ or
$|\downarrow\rangle\otimes|\uparrow\rangle$. The new state of the
system can be detailed as follows:
\begin{eqnarray}
|\Phi^-_{h}\rangle\!\!\!&=&\!\!\!\frac{1}{2\sqrt{2}}
\big[(-|R^{\uparrow}R^{\uparrow}\rangle
\!+\!|L^{\downarrow}L^{\downarrow}\rangle)_{ab}\!\otimes\!(|\uparrow\uparrow\rangle
\!-\!|\downarrow\downarrow\rangle)_{e_ae_b}\nonumber\\
&& +(|R^{\uparrow}L^{\downarrow}\rangle
\!-\!|L^{\downarrow}R^{\uparrow}\rangle)_{ab}
\!\otimes\!(|\uparrow\downarrow\rangle
\!-\!|\downarrow\uparrow\rangle)_{e_ae_b} \big].\;\;\;\;\;\;\;\;
\label{threet-outputS}
\end{eqnarray}
It can be viewed as a high-dimensional entanglement between the
photonic subsystem and electron-spin subsystem. After the parties
measure their photons, they can share two-QD entanglement. For
instance, if the outcome of the measurement on photons $ab$ is
$|R^{\uparrow}R^{\uparrow}\rangle_{ab}$ or
$|L^{\downarrow}L^{\downarrow}\rangle_{ab}$, the parties can get the
QD subsystem maximally entangled in the  state  as
\begin{eqnarray} 
|\Phi^-_2\rangle_{e}=\frac{1}{\sqrt{2}}(|\uparrow\uparrow\rangle
-|\downarrow\downarrow\rangle)_{e_ae_b}.
\label{threet-QDS}
\end{eqnarray}
However, if the  outcome of the measurement on
photons $ab$ is
$|R^{\uparrow}L^{\downarrow}\rangle_{ab}$ or
$|L^{\downarrow}R^{\uparrow}\rangle_{ab}$, an additional bit-flip operation
$\sigma_{x}^b=|\uparrow\rangle\langle\downarrow|+|\downarrow\rangle\langle\uparrow|$
on the electron $e_b$,  can also project the QD subsystem $e_ae_b$
 into the desired entangled state
$|\Phi^-_2\rangle_{e}$.

As for the faithful entanglement distribution of  the $N$-photon state, the parties can place
an entanglement source that generates $ N$ photons entangled in  GHZ
state
$|\Phi^+_{_N}\rangle_s=\frac{1}{\sqrt{2}}(|H\rangle_a|H\rangle_b\dots|H\rangle_z
+|V\rangle_a|V\rangle_b\dots|V\rangle_z)$
among the parties involved. With the similar encoder procedure to
that above, the parties can get their QDs entangled in the GHZ state
$|\Phi^-_{_N}\rangle_{e}=\frac{1}{\sqrt2}
(|\uparrow\uparrow\dots{}\uparrow\rangle
-|\downarrow\downarrow\dots{}\downarrow\rangle){e_ae_b\cdots{}e_z}$,
as shown in Appendix \ref{AppendixA}.

\subsection{Effective entanglement extension with efficient PCDs }
\label{sec2D}

\begin{figure}[!tp]
\begin{center}
\includegraphics[width=7.8 cm]{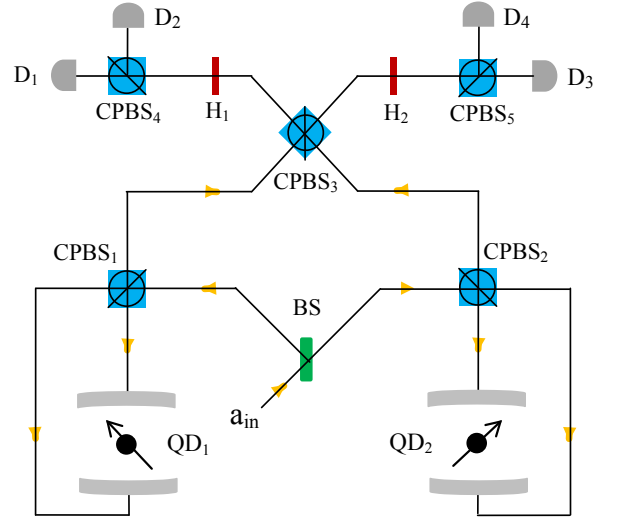}
\caption{(Color online) Schematic diagram of the efficient PCD on the two
QDs in the same node. H$_1$ and H$_2$ are two half-wave plates and
each is used to complete the Hadamard rotation on the circularly
polarized photons.} \label{fig-pcd}
\end{center}
\end{figure}

After the successful generation of the nonlocal $N$-electron GHZ state
$|\Phi^-_{_N}\rangle_{e}$ and several Bell states
$|\Phi^-_2\rangle_{e}=\frac{1}{\sqrt2}(|\uparrow\uparrow\rangle-|\downarrow\downarrow\rangle)$
for two QDs confined, respectively, in distant cavities separated
within the attenuation length, one can extend the length of the
quantum channel by local entanglement swapping, which can be
performed efficiently with our PCD shown in Fig. \ref{fig-pcd}.
Instead of subsequently inputting  one probe photon into two target
cavities \cite{Qdcoupling3,Qdqr1}, one can split
the incident photon into two spatial modes with a $50/50$ beam
splitter (BS), and then send each mode into one cavity, respectively. In
other words, only one effective input-output process  is involved in
our PCD, which makes it more efficient than others, especially in the
lower coupling regime.

Suppose the two stationary spin qubits $e_1$ and $e_2$ confined in $QD_1$ and $QD_2$ are in arbitrary
superposition states
$|\Phi\rangle_{e_1}=\alpha_1|\uparrow\rangle+\beta_1|\downarrow\rangle$
and
$|\Phi\rangle_{e_2}=\alpha_2|\uparrow\rangle+\beta_2|\downarrow\rangle$,
respectively.  Here $|\alpha_{1}|^2 + |\beta_{1}|^2=|\alpha_{2}|^2 +
|\beta_{2}|^2 = 1$.  A polarized photon \emph{p} in the state
$|\Phi\rangle_{p}=\frac{1}{\sqrt2}(|R\rangle+|L\rangle)$ input
into the \emph{a$_{in}$} import of the PCD (shown
in Fig. \ref{fig-pcd}). After passing through the BS, it is changed
into the state
\begin{eqnarray}
|\Phi_1\rangle_{p}=\frac{1}{2}(|R\rangle+|L\rangle)_a\otimes(|a_1\rangle+|a_2\rangle),
\end{eqnarray}
where $|a_1\rangle$ and $|a_2\rangle$ are two spatial modes of the
photon \emph{a} that are sent to $QD_1$ and $QD_2$, respectively.
The photon in the states $|R\rangle$ and $|L\rangle$ is separated by
CPBSs, and then enters the cavities. When the photon leaves the cavities, the
system composed of the photon \emph{p}, $e_1$, and $e_2$   evolves
into the state $|\Phi_h\rangle_{1}$. Here
\begin{eqnarray}
|\Phi_h\rangle_{1}\!\!&=&\!\!\frac{1}{2}(|R^{\uparrow}\rangle\!+\!|L^{\downarrow}\rangle)_a
\!\otimes\![|a_1\rangle\!\otimes\!(\alpha_1|\uparrow\rangle\!-\!\beta_1|\downarrow\rangle)_{e_1}
\!\otimes\!|\Phi\rangle_{e_2}\nonumber\\
&&\!+\;|\Phi\rangle_{e_1}\!\otimes\!|a_2\rangle\!\otimes\!(\alpha_2|\uparrow\rangle\!-\!\beta_2|\downarrow\rangle)_{e_2}].
\end{eqnarray}
CPBS$_1$ and CPBS$_2$ are used to combine the photon in the states
$|R^{\uparrow}\rangle$ and $|L^{\downarrow}\rangle$ in each spatial
mode.  The two spatial modes $|a_1\rangle$ and $|a_2\rangle$ of
photon \emph{p} interfere with each other at  CPBS$_3$, and
then a Hadamard rotation H$_1$ or H$_2$ on the photon \emph{p} is
applied. The state of the system composed of \emph{p}, $e_1$, and
$e_2$ becomes
\begin{eqnarray}
|\Phi_h\rangle_{2}\!&=&\!\frac{1}{\sqrt{2}}[|R_{a_1}\rangle
\otimes(\alpha_1\alpha_2|\uparrow,\uparrow\rangle-\beta_1\beta_2|\downarrow,\downarrow\rangle)_{e_1e_2}\!\nonumber\\
&& +\; |L_{a_1}\rangle\otimes(\beta_1\alpha_2|\downarrow,\uparrow\rangle-\alpha_1\beta_2|\uparrow,\downarrow\rangle)_{e_1e_2}\!\nonumber\\
&&+\; |R_{a_2}\rangle \otimes(\alpha_1\alpha_2|\uparrow,\uparrow\rangle-\beta_1\beta_2|\downarrow,\downarrow\rangle)_{e_1e_2}\!\nonumber\\
&&+\;
|L_{a_2}\rangle\otimes(\alpha_1\beta_2|\uparrow,\downarrow\rangle-\beta_1\alpha_2|\downarrow,\uparrow\rangle)_{e_1e_2}].
\;\;\;\;\;\;\;\;\;
\end{eqnarray}
Here, the subscripts ${a_1}$ and ${a_2}$ represent the spatial modes
of photon \emph{p} sent to the left analyzer and the right one,
respectively. One can project the state of the two spins
nondestructively into the state
\begin{eqnarray}
|\Phi^{_E}_2\rangle_{e}=\frac{1}{\sqrt{2}}
(\alpha_1\alpha_2|\uparrow,\uparrow\rangle-\beta_1\beta_2|\downarrow,\downarrow\rangle)_{e_1e_2},
\label{outputpcd}
\end{eqnarray}
when a $|R\rangle$ polarized photon is detected by the single-photon  detectors; otherwise, the state of the two spins will collapse into
\begin{eqnarray}
|\Phi^{_O}_2\rangle_{e}=\frac{1}{\sqrt{2}}
(\alpha_1\beta_2|\uparrow,\downarrow\rangle-\beta_1\alpha_2|\downarrow,\uparrow\rangle)_{e_1e_2}.
\label{outputpcd2}
\end{eqnarray}
That is to say, the click of the photon detector D$_1$ or
D$_3$ announces the even parity of the two spins and the click of
D$_2$ or D$_4$ heralds the odd parity of the two spins.

Considering  $N+1$ communication nodes, say, Alice ($e_a$), Bob
($e_b$),\dots, Zach ($e_z$) and Dean ($e_d$), the original
$N$-electron GHZ state shared by Alice, Bob,\dots, and Zach is
$|\Phi^-_{_N}\rangle_{e}=\frac{1}{\sqrt2}
(|\!\uparrow\uparrow\dots{}\uparrow\rangle
-|\!\downarrow\downarrow\dots{}\downarrow\rangle){e_ae_b\cdots{}e_z}$, and the Bell state shared by Zach
($e_{z'}$) and Dean is  $|\Phi^-_2\rangle_{e}=\frac{1}{\sqrt2}(|\!\uparrow\uparrow\rangle-|\!\downarrow\downarrow\rangle)$. The total spin
state of the  $N'=N+2$  electrons can be written as
\begin{eqnarray}
|\Phi_{h}\rangle_{3}\!&=&\!|\Phi^-_{_N}\rangle_{e}\otimes|\Phi^-_2\rangle_{e}\nonumber\\
\!&=&\!\frac{1}{2}(|\uparrow,\uparrow,\dots,\uparrow\rangle
-|\downarrow,\downarrow,\dots,\downarrow\rangle)_{e_a\cdots{}e_z}\nonumber\\
&&\otimes\;(|\uparrow,\uparrow\rangle-|\downarrow,\downarrow\rangle)_{e_{z'}e_d}.\;\;\;\;\;\;\;\;\;\;\;\;\;\;\;
\;\;\;\;\;\;\;\;\;\;\;\;\;
\end{eqnarray}
If Zach applies the PCD on the two QDs $e_{z}$ and
$e_{z'}$,  the  system composed of the  $N'$  electrons will evolve
into  $|\Phi_h\rangle_{4}$, before the click of photon detectors in
the PCD shown in Fig. \ref{fig-pcd}. Here
\begin{eqnarray}
|\Phi_h\rangle_{4}\!\!&=&\!\!\frac{1}{2\sqrt{2}}[(|R_{a_1}\rangle +|R_{a_2}\rangle) \otimes(|\uparrow,\cdots\uparrow, \uparrow,\uparrow\rangle \nonumber\\
&&-\;|\downarrow,\cdots,\downarrow,\downarrow,\downarrow\rangle)_{e_a\cdots{}e_z,e_{z'}e_d}\!\nonumber\\ && +\; (|L_{a_1}\rangle-|L_{a_2}\rangle)\otimes
(|\uparrow,\cdots\uparrow, \downarrow,\downarrow\rangle \nonumber\\
&&-\;|\downarrow,\cdots,\downarrow,\uparrow,\uparrow\rangle)_{e_a\cdots{}e_z,e_{z'}e_d}],\!
\;\;\;\;\;\;\;\;\;\;\;\;\;\;\;\;\;\;\;\;\;\;\;\;
\end{eqnarray}
where the $N'$ stationary QDs are divided into the even parity case
$|\Phi^E_{N'}\rangle_{e}$ and the odd-parity one
$|\Phi^O_{N'}\rangle_{e}$ conditioned on the detection of
$|R\rangle$ and $|L\rangle$ photon, respectively, i.e.,
\begin{eqnarray}\label{QSn2}
|\Phi^E_{N'}\rangle_{e}\!\!&=&\!\!\frac{1}{\sqrt{2}}(|\!\uparrow,\cdots,\uparrow, \uparrow,\uparrow\rangle
\!-\!|\!\downarrow,\cdots,\downarrow,\downarrow,\downarrow\rangle)_{e_a\cdots{}e_z,e_{z'}e_d}, \nonumber\\
|\Phi^O_{N'}\rangle_{e}\!\!&=&\!\!\frac{1}{\sqrt{2}}
(|\!\uparrow,\cdots,\uparrow,\downarrow,\downarrow\rangle \!-\!|\!\downarrow,\cdots,\downarrow,\uparrow,\uparrow\rangle)_{e_a\cdots{}e_z,e_{z'}e_d}. \nonumber\\
\end{eqnarray}
With these $N'$-spin GHZ states, Zack performs a Hadamard operation
on the two QDs $e_z$ and $e_{z'}$, and then he measures the states of
$e_z$ and $e_{z'}$ with the basis $\{|\uparrow\rangle,
|\downarrow\rangle\}$, which  will project the remaining $N$ QDs
$e_a$, $e_b$, $\cdots$, and $e_d$ into the desired GHZ state with the
form $|\Phi^-_{_N}\rangle_{e}$, up to a local operation on
{$e_d$}. The span of the GHZ quantum channel is eventually further
extended.  Meanwhile, if one takes another GHZ state instead of the
Bell state  $|\Phi^-_{2}\rangle_{e}$ to perform the entanglement
extension, the number of the parties involved in the repeater can
also be increased and the parties in the quantum communication network
can,  in principle, extend arbitrarily their communication distance with the same
quantum entanglement extension process described above.

\section{Entanglement purification to depress the influence of
asymmetric noise on different time bins}
\label{sec3}

We have detailed the principle and the process of our heralded
quantum repeater protocol, in which the influences of the noisy
channels on the early and the late time bins of the photons are
considered to be the same one and the fluctuation of the noise at
the nanosecond scale has been neglected. In a practical condition, maybe
the channel is  noisy and the fiber parameters have local fast
variations. The influences of the noisy channels can vary  at
the different time bins. The unitary transformation $U_i^l$ at the
late time bin $|l\rangle$ is different from   $U_i^s$ at the
early time bin $|s\rangle$. At this time, the final entangled state
of the two-QD subsystem  will be less entangled and the entanglement
purification process \cite{Qrepeater0,high-frep1,repeaterwangc} is required to obtain the maximally entangled
state for nonlocal electron-spin systems.

Suppose the influences of the noisy channels at the different time
bins are of  little difference. The unitary transformations on  the
early time bin and the late one are described with $U_i^s$ and
$U_i^l$, respectively,
\begin{eqnarray}
\begin{split}
U_i^s|H\rangle_i\;\;=\;\;&\delta_i|H\rangle_i+\eta_i|V\rangle_i, \\
U_i^l|H\rangle_i\;\;=\;\;&\delta'_i|H\rangle_i+\eta'_i|V\rangle_i,
\label{threet-initialAS}
\end{split}
\end{eqnarray}
where $|\delta_i|^2+|\eta_i|^2=|\delta'_i|^2+|\eta'_i|^2=1$ ($i=a$,
$b$ ). The state of the two-photon system after passing
through the noisy channels can be described as
\begin{eqnarray}
|\Phi^+_2\rangle_{t_4}\!\!\!&=&\!\!\!
\frac{1}{\sqrt{2}}\big[|s\rangle_a|s\rangle_b(\delta_a|H\rangle_a+\eta_a|V\rangle_a)
(\delta_b|H\rangle_b+\eta_b|V\rangle_b)\nonumber\\
&&\!\!\!+|l\rangle_a|l\rangle_b(\delta'_a|H\rangle_a+\eta'_a|V\rangle_a)
(\delta'_b|H\rangle_b+\eta'_b|V\rangle_b)\big],\;\;\;\nonumber\\
\label{threet-initial}
\end{eqnarray}
which is a partially entangled Bell state when considering the
time-bin qubits but the polarization states of the photons $ab$ at
different time bins are ambiguous and separable. With the same
decoder in each node shown in Fig. \ref{fig3} (b), the early
components of the photons $ab$ will be converted into the time-bin
qubits with the polarization state $|V\rangle_a|V\rangle_b$, while
the late components of $ab$ will be converted into another kind of
time-bin qubits with the polarization state $|H\rangle_a|H\rangle_b$
which equals to the early one only when the symmetric noise model
is effective ($U_i^s=U_i^l$). After Bob performs a $\pi$ phase shift on the $|V\rangle_b$ component of the photon $b$, and before the photon $b$ passes through
the QWPs placed near the imports of the
cavities,  the state of the photonic subsystem can be written as follows:
\begin{eqnarray}
|\Phi^+_2\rangle_{t_5}\!\!&=&\!\!\!
\frac{1}{\sqrt{2}}\big[-|V^{\downarrow}\rangle_a|V^{\downarrow}\rangle_b (\delta_a|s'\rangle_a+\eta_a|l'\rangle_a)\nonumber\\
&&\!\!\otimes(\delta_b|s'\rangle_b+\eta_b|l'\rangle_b)+|H^{\uparrow}\rangle_a|H^{\uparrow}\rangle_b\nonumber\\
&&\!\!
\otimes(\delta'_a|s'\rangle_a+\eta'_a|l'\rangle_a)(\delta'_b|s'\rangle_b+\eta'_b|l'\rangle_b)
\big],\;\;\;\;\;\;\;\;
\end{eqnarray}
which is a partially entangled polarization state when the time bin
information is determined. Since the unitary transformations are
arbitrary and unknown, one can describe the photonic state with the
 density matrix $\rho$,
\begin{eqnarray}
\rho=\mu|\Phi^-_2\rangle_0\langle\Phi^-_2|+(1-\mu) |\Phi^+_2\rangle_0\langle\Phi^+_2|,
\end{eqnarray}
which can be viewed as a mixture of
$|\Phi^{-}_2\rangle_0=\frac{1}{\sqrt{2}}(|H^{\uparrow}\rangle_a|H^{\uparrow}\rangle_b
-|V^{\downarrow}\rangle_a|V^{\downarrow}\rangle_b)$
and
$|\Phi^{+}_2\rangle_0=\frac{1}{\sqrt{2}}(|H^{\uparrow}\rangle_a|H^{\uparrow}\rangle_b
+|V^{\downarrow}\rangle_a|V^{\downarrow}\rangle_b)$
with the probabilities  $\mu$ and $1-\mu$, respectively.

We would like to consider first the case that the photons $ab$ are in the
state $|\Phi^{+}_2\rangle_0$ before entering the cavities. With a similar process to that for
$|\Phi^{-}_2\rangle_0$, we can complete the entanglement transfer
from the photonic subsystem to the QDs subsystem, since the relative phase
between different polarization modes of the photons will be mapped
into the relative phase between the different spin states of the
QD-confined electrons. The state of the hybrid system composed of
the photons $ab$ and the electron spins $e_ae_b$
after the interactions evolves to $|\Phi^+_h\rangle$, instead of
$|\Phi^-_h\rangle$ shown in Eq. (\ref{threet-outputS}). Here
\begin{eqnarray}
|\Phi^+_{h}\rangle\!\!\!&=&\!\!\!\frac{1}{2\sqrt{2}}
\big[(|R^{\uparrow}R^{\uparrow}\rangle
\!+\!|L^{\downarrow}L^{\downarrow}\rangle)_{ab}\!\otimes\!(|\downarrow\downarrow\rangle
\!+\!|\uparrow\uparrow\rangle)_{e_ae_b}\nonumber\\
&&+(|R^{\uparrow}L^{\downarrow}\rangle
\!+\!|L^{\downarrow}R^{\uparrow}\rangle)_{ab}
\!\otimes\!(|\downarrow\uparrow\rangle
\!+\!|\uparrow\downarrow\rangle)_{e_ae_b} \big].\;\;\;\;\;\;\;\;
\label{threet-outputpp}
\end{eqnarray}
Comparing the hybrid state  shown in  Eq. (\ref{threet-outputpp}) with that in Eq.
(\ref{threet-outputS}), one can easily see that when the
 input photons are in the mixed state $\rho$,  the detection of
one photon in each node will project the electron spins
$e_ae_b$ into another mixed state  $\rho''$ with or without an additional
single-qubit bit-flip operation on $e_a$,
\begin{eqnarray}
\rho'' = \mu|\Phi_2^-\rangle_{e_0}\langle\Phi_2^-| +
(1-\mu)|\Phi_2^+\rangle_{e_0}\langle\Phi_2^+|.
\end{eqnarray}
It is a mixture of two-QD Bell states $|\Phi_2^-\rangle_{e_0}=1/\sqrt{2}(|\uparrow\uparrow\rangle-|\downarrow\downarrow\rangle)$ and
$|\Phi_2^+\rangle_{e_0}=1/\sqrt{2}(|\uparrow\uparrow\rangle+|\downarrow\downarrow\rangle)$ with the probabilities $\mu$ and $1-\mu$,
respectively.

When the parties in the quantum communication network have obtained the mixed
state $\rho''$, they can use entanglement purification to increase
the fidelity of the entangled channel between Alice and Bob. Since
the phase-flip error cannot be purified directly, the parties can
perform a Hadamard operation on each QD and  convert the joint
state of $e_a$ and $e_b$  into
\begin{eqnarray} 
\rho''_h= \mu|\Phi_{0}^{'-}\rangle_e\langle\Phi_{0}^{'-}|
+ (1-\mu)|\Phi_{0}^{'+}\rangle_e\langle\Phi_{0}^{'+}|.
\end{eqnarray}
Here $|\Phi_{0}^{'-}\rangle_e\!
=\frac{1}{\sqrt{2}}(|\uparrow\downarrow\rangle+|\downarrow\uparrow\rangle)$ and $ |\Phi_{0}^{'+}\rangle_e=\frac{1}{\sqrt{2}}(|\uparrow\uparrow\rangle
+|\downarrow\downarrow\rangle)$.
The original phase-flip error is  mapped into a bit-flip error, and
Alice and Bob can perform the entanglement purification process with
our efficient PCD to improve the fidelity of the mixed state $\rho''_h$. Its principle can be described in detail as follows.

Alice and Bob can take two copies of QD systems $e_ae_b$ and
$e'_ae'_b$ for each round of purification  and each system is in the
state $\rho''_h$. The composite four-QD system is in the state
$\rho_{P}^{T}$ which could be viewed as the mixture of four pure
states
$|\Phi_{0}^{'-}\rangle_{e_ae_b}\otimes|\Phi_{0}^{'-}\rangle_{e_a^{'}e_b^{'}}$,
$|\Phi_{0}^{'-}\rangle_{e_ae_b}\otimes|\Phi_{0}^{'+}\rangle_{e_a^{'}e_b^{'}}$,
$|\Phi_{0}^{'+}\rangle_{e_ae_b}\otimes|\Phi_{0}^{'-}\rangle_{e_a^{'}e_b^{'}}$,
and
$|\Phi_{0}^{'+}\rangle_{e_ae_b}\otimes|\Phi_{0}^{'+}\rangle_{e_a^{'}e_b^{'}}$
with the probabilities of $\mu^2$, $\mu(1-\mu)$, $\mu(1-\mu)$, and
$(1-\mu)^2$, respectively. After the PCDs
performed by Alice and Bob, if all the outcomes are even,
the total four-QD system $e_ae_be'_ae'_b$  will be projected into
the state,
\begin{eqnarray}
|\varphi\rangle=\frac{1}{\sqrt{2}}(
|\uparrow\downarrow\uparrow\downarrow\rangle
\!+\!|\downarrow\uparrow\downarrow\uparrow\rangle)_{e_ae_be'_ae'_b},
\end{eqnarray}
with the probability of $\frac{\mu^2}{2}$  and
\begin{eqnarray}
|\varphi^{'}\rangle=\frac{1}{\sqrt{2}}(
|\uparrow\uparrow\uparrow\uparrow\rangle
\!+\!|\downarrow\downarrow\downarrow\downarrow\rangle)_{e_ae_be'_ae'_b},
\end{eqnarray}
with the  probability of $\frac{(1-\mu)^{2}}{2}$, respectively. If
both  Alice and Bob get an odd-parity result, they perform a
bit-flip operation on their  electron spins $e_a$ and $e_b$, which
leads to the same projection of the QD system as the case  that both outcomes
of the two PCDs are even. As for the case with one odd parity and
one even parity, which originates from the cross state
$|\Phi_{0}^{'+}\rangle_{e_ae_b}\otimes|\Phi_{0}^{'-}\rangle_{e_a^{'}e_b^{'}}$
and
$|\Phi_{0}^{'-}\rangle_{e_ae_b}\otimes|\Phi_{0}^{'+}\rangle_{e_a^{'}e_b^{'}}$,
it leads to the error and should be discarded. In other words, with
the PCDs, Alice and Bob can project the QD system
$e_ae_be_a^{'}e_b^{'}$  into
\begin{eqnarray}
\rho''_{h_1}\!=\!\frac{\mu^{2}}{\mu^{2}+(1-\mu)^{2}}|\varphi\rangle\langle\varphi|
\!+\!\frac{(1-\mu)^{2}}{\mu^{2}\!+\!(1-\mu)^{2}}|\varphi^{'}\rangle\langle\varphi^{'}|,\;\;\;\;
\end{eqnarray}
with the probability of ${\mu^{2}+(1-\mu)^{2}}$, when their outcomes
are the same ones in their  PCD processes.

In order to obtain the entangled state of a two-QD subsystem
$e'_ae'_b$, both Alice and Bob perform a Hadamard operation on their
electron spins $e_a$ and $e_b$. By measuring the spin states of the
QDs $e_a$ and $e_b$ with the basis $\sigma_z
=\{|\uparrow\rangle,\,|\downarrow\rangle\}$, they can, with or
without some phase-flip operations, get the desired QD subsystem
$e'_ae'_b$ in the states $|\Phi^{'-}_0\rangle_e$ and
$|\Phi^{'+}_0\rangle_e$ with the probabilities of
$\frac{\mu^{2}}{\mu^{2}+(1-\mu)^{2}}$ and
$\frac{(1-\mu)^{2}}{\mu^{2}+(1-\mu)^{2}}$, respectively. Finally,
another Hadamard operation on $e'_a$ and $e'_b$ will convert the
states $|\Phi^{'-}_0\rangle_e$ and $|\Phi^{'+}_0\rangle_e$ back into
$|\Phi_{2}^{-}\rangle_{e_0}$ and $|\Phi_{2}^{+}\rangle_{e_0}$,
respectively, leaving the whole system in the state,
\begin{eqnarray}
\rho_f'' \!=\!
\frac{\mu^{2}}{\mu^{2}\!+\!(1\!-\!\mu)^{2}}|\Phi_2^-\rangle_{e_0}\langle\Phi_2^-|
\!+\!
\frac{(1\!-\!\mu)^{2}}{\mu^{2}\!+\!(1\!-\!\mu)^{2}}|\Phi_2^+\rangle_{e_0}\langle\Phi_2^+|.\nonumber\\
\end{eqnarray}
It is a mixed entangled state with a higher fidelity than that in
the original one $\rho''$ when $\mu
>1/2$.

Certainly, the parties can improve further the fidelity of the
nonlocal quantum systems by iterating the purification protocol
several rounds with the method described above. For instance, if the
initial state with the fidelity $\mu > 0.7$ is available, they can
achieve the state with the fidelity $F> 0.997$ for only two rounds.

\section{ Influence on fidelity and efficiency from the practical
circular birefringence} \label{sec4}

In the discussion above, the spin-selection rule is taken to be
perfect and $2g^2/\kappa\gamma\gg1$, and the resonant
condition $|\Delta|\simeq 0$ is satisfied. In fact, the heavy-light
hole mixing can reduce the fidelity of the optical selection rules
\cite{hole-mixing}, and it can be improved for charged excitons due
to the quenched exchanged interaction \cite{hole-mixings}.
Meanwhile, the finite linewidth of the input light pulse will
inevitably make the resonant condition diffusion. The side leakage
of the cavity $\kappa_s$ and the limited coupling strength $g$ will
lead to the imperfect birefringent propagation of the input
photons \cite{Qdcoupling3,Qdcoupling4} as well. For instance, when the
electron spin is in the spin-up state $|\uparrow\rangle$, the
incident photon $|R^{\uparrow}\rangle$ or $|L^{\downarrow}\rangle$
totally reflected in the ideal case has a probability $t$ to be
transmitted through the cavity, and $|L^{\uparrow}\rangle$ or
$|R^{\downarrow}\rangle$ supposed to be totally transmitted has a
probability $r_0$ to be reflected.

To discuss the influence of the imperfect circular birefringence for
the QD-cavity unit on the fidelity of the quantum distribution
process, let us take the entanglement distribution with the
symmetric noise model as an example. In this case, the photons  $a$
and $b$ input into the cavities are in the state
$|\Phi^-_{2}\rangle_c=(|R^{\uparrow},R^{\uparrow}\rangle-|L^{\downarrow},L^{\downarrow})/\sqrt{2}$,
and the QD-confined electron spins $e_i$ ($i=a,b$) are all
initialized to  the  state
$|\Phi\rangle_{e_i}=\frac{1}{\sqrt2}(|\uparrow\rangle+|\downarrow\rangle)$.
The  simplified transformation relationship described in Eqs.
(\ref{transup}) and  (\ref{transdown}) for an ideal QD-cavity unit
should be modified, and the original transformation in Eq.
(\ref{hotrt1}) becomes dominant. When the electron spin is in the
spin-up state $|\uparrow\rangle$, one has the following
transformations:
\begin{eqnarray}
\begin{split}
|R^{\uparrow},\uparrow\rangle \;\; \rightarrow \;\;&
r|L^{\downarrow},\uparrow\rangle+t|R^{\uparrow},\uparrow\rangle,  \\
|R^{\downarrow},\uparrow\rangle  \;\; \rightarrow  \;\;&  t_0|R^{\downarrow},\uparrow\rangle+r_0|L^{\uparrow},\uparrow\rangle, \\
|L^{\downarrow},\uparrow\rangle   \;\; \rightarrow  \;\;&
r|R^{\uparrow},\uparrow\rangle+t|L^{\downarrow},\uparrow\rangle, \\
|L^{\uparrow},\uparrow\rangle  \;\; \rightarrow  \;\;&
t_0|L^{\uparrow},\uparrow\rangle+r_0|R^{\downarrow},\uparrow\rangle,
\label{transuppp}
\end{split}
\end{eqnarray}
where $r$ ($t$ ) and $r_0$ ($t_0$) are the reflection (transmission)
coefficients shown in Eqs. (\ref{hotrt}) and  (\ref{coldrt}),
respectively.  In other words,  the incident circularly polarized
photon $|R^{\uparrow}\rangle$ or $|L^{\downarrow}\rangle$ totally
reflected in the ideal case has a probability $t$ to be transmitted
through the cavity, and  $|L^{\uparrow}\rangle$ or
$|R^{\downarrow}\rangle$ supposed to be totally transmitted has a
probability $r_0$ to be reflected. When the excess electron is in
the state $|\downarrow\rangle$, the evolution can be described
similarly as:
\begin{eqnarray}
\begin{split}
|R^{\uparrow},\downarrow\rangle  \;\; \rightarrow  \;\;&
t_0|R^{\uparrow},\downarrow\rangle+r_0|L^{\downarrow},\downarrow\rangle, \\
|R^{\downarrow},\downarrow\rangle  \;\; \rightarrow  \;\;&
r|L^{\uparrow},\downarrow\rangle+t|R^{\downarrow},\downarrow\rangle, \\
|L^{\uparrow},\downarrow\rangle  \;\; \rightarrow  \;\;&
r|R^{\downarrow},\downarrow\rangle+t|L^{\uparrow},\downarrow\rangle, \\
|L^{\downarrow},\downarrow\rangle  \;\; \rightarrow  \;\;&
t_0|L^{\downarrow},\downarrow\rangle+r_0|R^{\uparrow},\downarrow\rangle.
\label{transdownpp}
\end{split}
\end{eqnarray}

According to the practical transformations in Eqs. (\ref{transuppp})
and  (\ref{transdownpp}), the non-normalized state of the composite hybrid
photon-QD system after the reflection of the photons $ab$ can be
written as
\begin{eqnarray}
|\Phi^{-'}_{h}\rangle\!\!&=&\!\!\!\frac{1}{2\sqrt{2}} \Big\{(|R^{\uparrow},R^{\uparrow}\rangle-|L^{\downarrow},L^{\downarrow}\rangle)
\big[(t^2-r^2)|\uparrow\uparrow\rangle\nonumber\\
&&+(r_0^2-t_0^2)|\downarrow\downarrow\rangle
+(tt_0-rr_0)(|\uparrow\downarrow\rangle+|\downarrow\uparrow\rangle)
\big]\nonumber\\
&&+
(rt_0-tr_0)(|R^{\uparrow},L^{\downarrow}\rangle-|L^{\downarrow},R^{\uparrow}\rangle)
\nonumber\\
&&\otimes(|\uparrow\downarrow\rangle+|\downarrow\uparrow\rangle)
\Big\}. \label{pradistr}
\end{eqnarray}
Conditioned on the click of one single-photon detector at each side,
the entanglement distribution process is supposed to be completed,
and the electron spins $e_ae_b$ will be collapsed into two different
partially entangled states  depending  on the outcomes of photon
detections.
\bigskip

To be detailed, when the outcomes of the measurements on photons $ab$
result in the even-parity space
$S=\{|R^{\uparrow},R^{\uparrow}\rangle,|L^{\downarrow},L^{\downarrow}\rangle\}$,
the electron spins $e_ae_b$ will be projected into the following
state:
\begin{eqnarray}
|\Phi^{-'}_{2}\rangle_e\!\!&=&\!\!\!\frac{1}{2\sqrt{2}}\big[(t^2-r^2)|\uparrow\uparrow\rangle
+(r_0^2-t_0^2)|\downarrow\downarrow\rangle\nonumber\\
&&+(tt_0-rr_0)(|\uparrow\downarrow\rangle+|\downarrow\uparrow\rangle)
\big], \label{pradistrrr}
\end{eqnarray}
with a success probability,
\begin{eqnarray}
\eta_d^{_E}=\frac{|2t+1|^2+|2t_0+1|^2+2|1+t+t_0|^2}{4}.
\end{eqnarray}
Note  $r\equiv1+t$ and $r_0\equiv1+t_0$ as shown in Eqs.
(\ref{hotrt}) and    (\ref{coldrt}). Therefore, the fidelity
$F^{_E}_d$ of the heralded entanglement between the  electron spins
$e_ae_b$ compared with the ideal target state
$|\Phi^{-}_2\rangle_{e}$ of entanglement distribution obtained with
the perfect birefringence in this case, can be detailed as
\begin{eqnarray}
F^{_E}_d&=&\big|{}_{e}\langle\Phi^{-}_2|\Phi^{-'}_{2}\rangle_e\big|^2\nonumber\\
   &=&\frac{|t^2-r^2-t_0^2+r_0^2|^2}{2(|t^2-r^2|^2+|r_0^2-t_0^2|^2+2|tt_0-rr_0|^2)}\nonumber\\
   &=&\frac{|t_0-t|^2}{2\eta_d^{_E}}.
\end{eqnarray}
In the other case, when the outcomes of photon detections belong to
the odd-parity space
$AS=\{|R^{\uparrow},L^{\downarrow}\rangle,|L^{\downarrow},R^{\uparrow}\rangle\}$,
the hybrid photon-QD in state $|\Phi^{-'}_{h}\rangle$ collapses the
electron spins $e_ae_b$ into the  state,
\begin{eqnarray}
|\Phi^{-'}_2\rangle_a&=&\frac{1}{\sqrt{2}}(|\uparrow\downarrow\rangle-|\downarrow\uparrow\rangle),
\label{polished-out}
\end{eqnarray}
with a probability of
\begin{eqnarray}
\eta_d^{_O}=\frac{|t_0-t|^2}{2}.
\end{eqnarray}
After Bob performs the bit-flip operation
$\sigma^b_{x}=|\uparrow\rangle\langle\downarrow|+|\downarrow\rangle\langle\uparrow|$
on the electron spin $e_b$, the same as  the case that the perfect circular
birefringence is effective, the state of the two electrons $e_ae_b$ evolves into
\begin{eqnarray}
|\Phi^{-'}_2\rangle_b&=&\frac{1}{\sqrt{2}}
(|\uparrow\uparrow\rangle-|\downarrow\downarrow\rangle),
\label{polished-out2}
\end{eqnarray}
which is the practical state of the  electron spins  $e_ae_b$ after the entanglement distribution,
 and it is identical to the target state $|\Phi^{-}_2\rangle_{e}$.
In other words, the fidelity $F^{_O}_{d}$  for this case is unity,
\begin{eqnarray}
F^{_O}_d=1,
\end{eqnarray}
which is independent of both the coupling strength $g/\kappa$ and
the cavity leakage $\kappa_s/\kappa$, and the corresponding
efficiency equals to $\eta^{_O}_d$.

\begin{figure}[!t]
\begin{center}
\includegraphics[width=8.5cm]{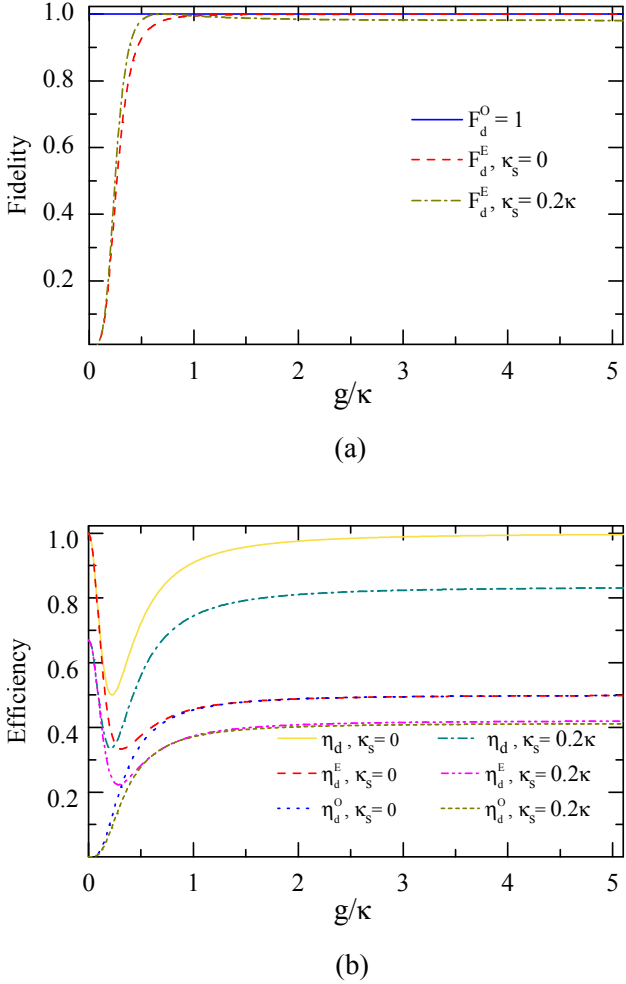}
\caption{(Color online) The performance of entanglement
distribution.  (a) The fidelities $F^O_d$ and $F^E_{d}$ conditioned
on different outcomes of photon detection in the entanglement
distribution  \emph{vs}  the normalized coupling strength
$g/\kappa$.  (b) The efficiency $\eta_d$  of  entanglement
distribution   \emph{vs} the normalized coupling strength
$g/\kappa$. Here  $\eta_d=\eta^{_O}_d+\eta^{_E}_d$,
${\gamma}/{\kappa}=0.1$, and the resonant condition
$\omega_c=\omega_{X^-}=\omega_0$ is adopted.} \label{Figs}
\end{center}
\end{figure}

From the discussion above, one can see that after entanglement
distribution the state of the electron spins $e_ae_b$ depends on the
outcomes of the photon detection  as a result of the imperfect
birefringent propagation of photons. That is,
$|\Phi^{-'}_2\rangle_e$ and $|\Phi^{-'}_2\rangle_b$ are conditioned
on the outcomes of photon detection in $S$ and $AS$, respectively.
Therefore, the total efficiency of the entanglement distribution can
be detailed as
\begin{eqnarray}
\eta_d=\eta_d^{_O}+\eta_d^{_E}=\frac{{|2t+1|^2+|2t_0+1|^2}}{2},
\end{eqnarray}
shown in Fig. \ref{Figs} (a). Meanwhile, the fidelities $F^{_E}_d$
and $F^{_O}_d$  measure  the overlap between the ideal target state
$|\Phi^{-}_2\rangle_{e}$ and  the practical  states
$|\Phi^{-'}_2\rangle_e$ and that between $|\Phi^{-}_2\rangle_{e}$
and $|\Phi^{-'}_2\rangle_b$, respectively. When the input maximally
entangled state of $ab$  is
$|\Phi_2^-\rangle_c=1/\sqrt{2}(|R^{\uparrow},R^{\uparrow}\rangle-|L^{\downarrow},L^{\downarrow})_{ab}$,
the ideal  target state $|\Phi^{-}_2\rangle_{e}$ of  $e_ae_b$ is
fully orthogonal to the original state of the two-electron system
that equals the practical one when spins $e_ae_b$ does not interact
with the photons, which leads to the vanish fidelity $F^{_E}_d=0$
for the outcome of the photon detection in  $S$ when the coupling
strength $g/\kappa=0$. However, when the outcome of the photon
detection  is in $AS$, we can get the unity fidelity $F^{_O}_{d}$
for entanglement distribution, even with the imperfect input-output
process, since the imperfect birefringence of the QD-cavity system
will appear as a whole coefficient; see Eq. (\ref{pradistrrr}) for
detail. However, for $g=0$, the corresponding efficiency
$\eta_d^{_O}$ for the outcome in $AS$  of the photon detection in
entanglement distribution vanishes $(\eta_d^{_O}=0)$, shown in Fig.
\ref{Figs}(b).

During the entanglement extension process, the party utilizes a PCD on two local
 spin qubits, and performs the partial measurements on the spins.
The influence of the practical circular birefringence on the
entanglement extension process can be estimated by the performance
of the PCD when the two QDs  are in the state
$|\Phi\rangle_{ee'}=\frac{1}{2}(|\uparrow\uparrow\rangle+
|\downarrow\uparrow\rangle+|\uparrow\downarrow\rangle+|\downarrow\downarrow\rangle)_{ee'}$.
The state of the composite hybrid system composed of the input probe photon
\emph{p} and the electron spin $ee'$  before the detection on the photon \emph{p} evolves
into
\begin{eqnarray}
|\Phi_h'\rangle_{1}\!\!&=&\!\!
\frac{1}{4\sqrt{2}}\Big\{(|R_1\rangle+|R_2\rangle)[2(r+t)|\uparrow\uparrow\rangle +2(r_0+t_0)\nonumber\\
&&\times|\downarrow\downarrow\rangle
+(r+t+r_0+t_0)(|\uparrow\downarrow\rangle+|\downarrow\uparrow\rangle) ]
\nonumber\\
&&+(|L_2\rangle-|L_1\rangle)(r+t-r_0-t_0)(|\uparrow\downarrow\rangle-|\downarrow\uparrow\rangle)\Big\},\nonumber\\
\label{praext}
\end{eqnarray}
and the corresponding success probability  $\eta_p$ can be detailed as
\begin{eqnarray}
\eta_p=\eta_p^{_O}+\eta_p^{_E}=\frac{|2t+1|^2+|2t_0+1|^2}{2},
\end{eqnarray}
which is identical to that for entanglement distribution $\eta_d$,
and it equals to the efficiency of the entanglement extension when
the single-qubit operation and the detection of the QD electron
spins are perfect
\cite{qqtime4,opticalpum1,opticalpum2,rotation,time1}. Here,
$\eta_p^{_O}=\eta_d^{_O}=(|t-t_0|^2)/2$ and
$\eta_p^{_E}=\eta_d^{_E}=({|2t+1|^2+|2t_0+1|^2+2|1+t+t_0|^2})/{4}$
represent the probabilities for the heralded success of the
entanglement extension by detecting an $|L\rangle$ (odd parity of
the PCD for two spins in the ideal case) and a $|R\rangle$ (even parity
of the PCD for two spins in the ideal case) polarized photons,
respectively. When $g/\kappa=0$, it leads to $\eta_p^{_O}=0$ no
matter whether $\kappa_s=0$  or not, since the probe photon  in
state $|\Phi\rangle_p=(|R\rangle+|L\rangle)/\sqrt{2}$ does not
interact with the QDs and it is still  in the state $|\Phi\rangle_p$
when it interferes with itself at  CPBS$_3$, shown in Fig.
\ref{fig-pcd}. This photon will be transformed into the $|R\rangle$
polarized photon by the Hadamard operation $H_1$ or $H_2$ and it
never leads to the click of detectors  $D_2$ or $D_4$, shown in Fig.
\ref{fig-pcd}.

To be detailed, when a $|R\rangle$ polarized photon is detected either
in the $|R_1\rangle$ or $|R_2\rangle$ mode, the QD subsystem $ee'$ will be collapsed
into
\begin{eqnarray}
|\Phi^{_{E'}}_2\rangle_{e}\!\!&=&\!\!
\frac{1}{2\sqrt{2}}
\Big[2(r+t)|\uparrow\uparrow\rangle +2(r_0+t_0)|\downarrow\downarrow\rangle\nonumber\\
&&+(r+t+r_0+t_0)(|\uparrow\downarrow\rangle+|\downarrow\uparrow\rangle)\Big].\;\;\;\;\;\;\;
\end{eqnarray}
The fidelity $F^{_E}_{p}$ of the PCD with respect to  the ideal
even-parity output  state
$|\Phi_2^{_E}\rangle_{e}=\frac{1}{\sqrt{2}}(|\uparrow\uparrow\rangle-|\downarrow\downarrow\rangle)_{ee'}$,
obtained from  Eq. \eqref{outputpcd} for
$\alpha_1=\alpha_2=\beta_1=\beta_2=1/\sqrt{2}$, can be detailed as
\begin{eqnarray}
F^{_E}_{p}&=&\big|{}_{e}\langle\Phi_2^{_E}|\Phi_2^{_{E'}}\rangle_e\big|^2\nonumber\\
&=&\frac{|r+t-r_0-t_0|^2}{2|r+t|^2+|r+t+r_0+t_0|^2+2|r_0+t_0|^2}\nonumber\\
     &=&\frac{|t_0-t|^2}{\eta^{_E}_{p}},
\end{eqnarray}
which is identical to the  fidelity $F^{_E}_{d}$ with the
even-parity outcome during the entanglement distribution process.
For $g/\kappa=0$, the QDs $ee'$ will be kept in the initial state
$|\Phi\rangle_{ee'}=1/2(|\uparrow\uparrow\rangle+|\downarrow\uparrow\rangle
+|\uparrow\downarrow\rangle+|\downarrow\downarrow\rangle)$, which is
fully orthogonal to the ideal output state
$|\Phi_0\rangle_{ee'}=1/\sqrt{2}(|\uparrow\uparrow\rangle-|\downarrow\downarrow\rangle)$,
leading to $F^{_E}_{p}=0$. In contrast, when an $|L\rangle$
polarized photon is detected, the QD subsystem $ee'$ will be
projected into the  state,
\begin{eqnarray}
|\Phi^{_{O'}}_2\rangle_{e}\!\!&=&\!\!
\frac{1}{\sqrt{2}}\big(|\uparrow\downarrow\rangle-|\downarrow\uparrow\rangle\big),
 \label{displ}
\end{eqnarray}
which is identical to the ideal odd-parity output  state
$|\Phi_2^{_{O}}\rangle_{e}=\frac{1}{\sqrt{2}}
(|\uparrow\downarrow\rangle-|\downarrow\uparrow\rangle)_{ee'}$
obtained from  Eq. \eqref{outputpcd2} with perfect birefringent
propagation when applying our PCD. Now, the fidelity $F^{_O}_{p}$ of
the state $|\Phi^{_{O'}}_2\rangle_{e}$  can be written as follows:
\begin{eqnarray}
F^{_O}_{p}=1.
 \label{extp}
\end{eqnarray}
The reason is that the photon detection and the interference of the
probe photon from different pathes transform the imperfect
birefringence into a nonlocal coefficient;  see Eq. (\ref{praext})
for detail.  When $g/\kappa=0$, the nonlocal coefficient will
vanish, which  results in the fact that the corresponding
probability $\eta_p^{_O}=0$.

The fidelities $F^{_O}_d$ and $F^{_E}_{d}$ conditioned on different
outcomes of the photon detection in the entanglement distribution
are shown in Fig. \ref{Figs} (a)  as the function of the side
leakage of the spin-cavity system $\kappa_s/\kappa$ and the coupling
strength $g/\kappa$ on the resonant interaction condition, where
${\gamma/\kappa}=0.1$ and $\omega_c=\omega_{X^-}=\omega$. Meanwhile,
the fidelities $F^{_O}_p$ and $F^{_E}_{p}$ in entanglement extension
are identical to  $F^{_O}_d$ and $F^{_E}_{d}$, respectively. When
the coupling strength  $g/\kappa >0.6$ , both the entanglement
distribution process and the entanglement extension process  are
near perfect, with the minimal fidelity $F^{_E}_d=F^{_E}_p >0.948$
for both cases with the side leakage $\kappa_s/\kappa=0$ and
$\kappa_s/\kappa=0.2$. When the coupling strength   $g/\kappa =1.2$,
the minimum fidelities of the entanglement distribution  and
 extension with the side leakage
$\kappa_s/\kappa=0.2$ are $F^{_E}_d=F^{_E}_p=0.991$. They can be
increased to be $F^{_E}_d=F^{_E}_p=0.998$ when $\kappa_s/\kappa=0$ and $g/\kappa =1.2$.
Interestingly, the detection of an $|L\rangle$ polarized photon in
entanglement extension and that of
$|R^{\uparrow}R^{\uparrow}\rangle$ or
$|L^{\downarrow}L^{\downarrow}\rangle$ in entanglement distribution
can lead to the corresponding error-free processes, no matter what
the coupling strength  $g/\kappa $ and the side leakage
$\kappa_s/\kappa$  are, and it is useful for generating entangled
states and scalable one-way quantum computation \cite{qqentangle1}.

The total efficiency  of   entanglement distribution
$\eta_d=\eta^{_O}_d +\eta^{_E}_d$ is  shown in Fig. \ref{Figs} (b)
with the same parameters as those for the fidelities. Meanwhile, the
efficiencies $\eta^{_O}_p$ and $\eta^{_E}_p$  in entanglement
extension are identical to the corresponding ones $\eta^{_O}_d$ and
$\eta^{_E}_d$ in entanglement distribution. One should note that
when the coupling rate $g$ is small, especially, $g/\kappa<0.6$, the
fidelities $F_d^{_E}$ and $F_p^{_E}$ are much smaller than
$F_d^{_O}=F_p^{_O}=1$, and we should treat these two kinds of
outcomes in entanglement distribution ($AS$ or $S$) and entanglement
extension ($|R\rangle$ or $|L\rangle$) independently. However, when
$g$ is large, the total efficiencies are more important, since both
kinds of outcomes are faithful, shown in Fig. \ref{Figs} (a). When
the coupling strength $g/\kappa=1.2$, the efficiencies
$\eta_d=\eta_p=0.770$ for the side leakage $\kappa_s/\kappa =0.2$.
If $g/\kappa=2.4$ and $\kappa_s/\kappa=0$, the efficiencies
$\eta_d=\eta_p=0.983$ are achievable. The small reduction from unity
probability  for $\kappa_s=0$ originates from the noise operator
$(\hat{N})$ associated with the spontaneous decay of the trion
state. Meanwhile, when we increase the side leakage to
$\kappa_s=0.2\kappa$, both $\eta_d^{_E}$ ($\eta_p^{_E}$) and
$\eta_d^{_O}$ ($\eta_p^{_O}$) decrease a little leading to the
decrease  in the total efficiency $\eta_d$ ($\eta_p$), due to the
finite reflection originating from the coupling to side leakage mode
$\hat{s}_{in}$. The efficiency $\eta_d$ ($\eta_p$)   decreases
scince the  increase of the side leakage  $\kappa_s/\kappa$ will
decrease the radiation into the cavity, resulting in a decrease of
the output photon in the subspace spanned by the transmission and
reflection modes, shown in Fig. \ref{fig2} (b). In addition, when the
achievable input-coupling efficiency $\eta_{in}=90\%$ is considered
\cite{q215}, the efficiencies above should be further reduced by
$19\%$ and $10\%$ for the efficiencies $\eta_d$ and $\eta_p$, since
two input-output processes is involved in the entanglement
distribution process while only one is involved in the entanglement extension
process with our efficient  PCD.

\section{Discussion and summary}
\label{sec5}

Thus far, we have detailed the process of establishing the quantum
entangled channel for the quantum communication network.  The photons
entangled in the time-bin DOF are exploited to entangle the remotely
separated QD-cavity units. Currently,  the  sources producing photon
pairs  with  polarization entanglement are well developed. With some
optical elements, the polarized entanglement can be transformed into
the time-bin one before the transmission over noisy optical-fiber
channels, shown in Fig. \ref{fig3}. Along with our effective PCD,
the parties  can perform the heralded extension of the entanglement
across the quantum network with quantum swapping, and increase the
entanglement with entanglement purification. In addition, by picking
out the outcome of the PCD in which an $|L\rangle$ polarized photon
is detected for the success signal, the two QDs will be projected
into the odd-parity  state in a heralded way and the influence of
imperfect birefringence on the entanglement purification and the
entanglement extension processes can be eliminated.

In our protocol, the electron spin of the QD  acts  as  a  quantum
node. Before the arrival of the incident photon,  the users
initialize their spins by optical pumping or optical cooling
\cite{opticalpum1,opticalpum2}, followed by single-spin rotations
\cite{rotation,qqtime4}. The time needed for the coherent control of
electron spins has been suppressed into the scale of picosecond in the
semiconductor quantum dot \cite{time1}. Meanwhile, an electron-spin
coherence time as high as $T_2\simeq 2.6$ $\mu{}$s has been
experimentally achieved \cite{timec}, which is quite long compared
with  preparation and measurement time (ns scales). Hence, the
cavity photon time $\tau=4.5$ ns will be the dominant time interval
for the exciton dephasing \cite{Qdcoupling4}. The previous fidelity
of the final quantum networking will be reduced by the amount of
$[1-exp(-\tau/T_2)\simeq0.002]$. When the absence of  nuclear spins
is achieved,  e.g., by using isotopically purified II-VI materials,
the decoherence time is theoretically predicted to be as long as the
spin relaxation time which is currently  $20$ ms at a magnetic field
$4T$ and at $1K$   \cite{qqtime2} and can be much longer for a lower
magnetic field \cite{qqtime4,qqtime3}.

Our scheme prefers the  strong coupling between the QDs and the
cavity, and it can also be performed with  low-$Q$-factor cavities
where $g/\kappa <1$
 at the price of decreasing the efficiency a bit. The strong coupling has
been observed in various QD-cavity systems \cite{strongp,strongcry}.
For micropillars with the diameter $d_c$ around 1.5 $\mu{}$m, the
$X^-$ dipole decay rate $\gamma/2\simeq1$ $\mu{}$eV  when the
temperature $T=2K$ \cite{QD-tem}. The coupling strength
$g=80$ $\mu{}$eV and the cavity quality factor including the side
leakage as high as $Q>4\times10^4$  has been experimentally realized
with $In_{0.6}Ga_{0.4}As$ in a similar experiment setup \cite{q4}.
In other words, $g/(\kappa+\kappa_s)>2.4$ is achievable. Meanwhile,
the coupling strength $g$ depends on the QD exciton oscillator
strength and   the mode volume $V$, while   $\kappa$ is
determined by the cavity quality factor, and
they can, in principle, be controlled independently to achieve a larger
$g/(\kappa+\kappa_s)$. Recently, the coupling strength $g=16
$ $\mu{}$eV and a cavity spectral width as low as $\kappa=20.5
$ $\mu{}$eV ($Q$=65 000) have been achieved in a  7.3 $\mu{}$m diameter
micropillar \cite{q65}. And then, the quality factor is improved to
$Q=2.15\times10^5$($\kappa=6.2$ $\mu{}$eV) with lower side leakage
\cite{q215}.

The imperfection that comes from photon loss is also an inevitable
problem in the previous schemes
\cite{high-frep1,high-frep2,revqr,ionqr,Qdqr1,Qdqr2}. The photon
loss occurs due to the cavity imperfection, the fiber absorption,
and the inefficiency of the single-photon detector. As the
successful generation of the electron-spin entangled state and the
completion of quantum extension are heralded by the detection of
photons, the photon loss will only affect the efficiency of our
scheme and has no effect on the fidelity of the quantum channel
established. During the transmission of the photons, there is no
restriction on the electron spins. That is,  if the photons can
arrive at the nodes, the distance of the adjacent nodes can be long,
different from those limited by the coherent time of the quantum
nodes \cite{Qrepeater0, revqr} as the entanglement between
neighboring nodes are constructed by entanglement swapping between
the stationary qubit and the flying qubit in the latter. The
efficiency of our entanglement distribution protocol is at least two
times  more than those performed with two-photon coincidence
detection \cite{Qdqr2,qqentangle2}, since the photonic  entanglement
can be totally converted into the QD entanglement conditioned on the
detection of one photon at each node. Meanwhile, the multi-mode
speed-up procedure \cite{QRmulti-spatial}  agrees with our protocol
and can be involved in a similar way to that presented by Jones
\emph{et al} \cite{qqentangle2}.

In summary, we have proposed an efficient quantum repeater protocol
for spin-photon systems with the help of the  time-bin encoder and the
generalized interface between the circularly polarized photon and
the QD embedded in a double-sided optical microcavity. It works in a
heralded  way and requires only one channel, not two or more
\cite{Qdqr3}. The users can establish a maximally entangled quantum
channel which is independent of the particular parameters of the
collective-noise channel. We also construct an efficient PCD based
on one effective input-output process of a single photon, and it can
simplify  the entanglement channel extension and entanglement
purification that is used to suppress the  phase-flip errors
originating from the imperfection of the collective-noise channel.
This protocol is  feasible with current technology and can find its
application directly in the quantum communication network protocols.

\section*{ACKNOWLEDGMENTS}

This work is supported by the National Natural Science Foundation of
China under Grants No. 11174039,  No. 11174040, and No. 11474026,  and the
Fundamental Research Funds for the Central Universities under Grant
No. 2015KJJCA01.



\appendix
\section{ $N$-user GHZ state distribution for a multiuser quantum repeater network}
\label{AppendixA}

The principle of our deterministic entanglement creation for two
legitimate participants can be extended to the $N$-participant case
directly. Assume the original local $N$-photon GHZ state in the
polarization DOF can be described as
\begin{eqnarray}
|\Phi^+_N\rangle=\frac{1}{\sqrt{2}}(|H\rangle_a|H\rangle_b\dots|H\rangle_z+|V\rangle_a|V\rangle_b\dots|V\rangle_z),\;\;\;
\label{N-initial0}
\end{eqnarray}
where the subscripts $a$, $b$, $\dots$, and $z$ represent the
photons directed to Alice, Bob, $\dots$, and Zach, respectively.
After  a similar encoder to that in the two-photon case performed  on each of
the $N$ photons, the state of  the system composed of the $N$ photons
$ab\dots{}z$ launched into the noisy quantum channels becomes
\begin{eqnarray}
|\Phi^+_N\rangle_{t_0}\!\!&=&\!\!\frac{1}{\sqrt{2}}|H\rangle_a|H\rangle_b\dots|H\rangle_z
\otimes(|s\rangle_a|s\rangle_b\dots|s\rangle_z\nonumber\\
&&+|l\rangle_a|l\rangle_b\dots|l\rangle_z).\;\;\;\;\;\;\;\;\;\;\;\;\;\;\;\;\;\;\;\;\;\;\;\;\;\;\;\;\;\;\;\;\;\;\;\;\;\;\;\;
\label{N-initial}
\end{eqnarray}
It is an $N$-qubit time-bin entanglement. Here $|s\rangle_i$ and
$|l\rangle_i$ ($i=$ $a$, $b$, $\dots$,  $z$) denote the components
of the  photon $i$ which pass through the short path and the long
path of the encoder shown in Fig. \ref{fig3} (a), respectively. To
address the influences of the collective-noise channels on the $N$
photons, we can introduce $N$ unknown unitary operators:
\begin{eqnarray}
U_i=\delta_i|H\rangle_i+\eta_i|V\rangle_i,
\end{eqnarray}
where the subscripts $i=a,b,\cdots,z$ are used to denote the noise operators acting on the photons $a$, $b$, $\dots$, and $z$, respectively.
Since the time separation between $|s\rangle_i$ and $|l\rangle_i$
time bins are of the nanosecond scale and taken to be much less than
the time of the noise fluctuation of the channels, the influence on
the $|s\rangle_i$ components is identical to that  on the
$|l\rangle_i$ component. After passing through the noisy channels, with a $\pi$ phase shift on one $|V\rangle$ polarized photon, i.e.,  $|V\rangle_a$,
the state of the system composed of the $N$ photons evolves into
\begin{eqnarray}
|\Phi^-_N\rangle_{t_1}\!&=&\!\frac{1}{\sqrt{2}}(|s\rangle_a|s\rangle_b\cdots|s\rangle_z
+|l\rangle_a|l\rangle_b\cdots|l\rangle_z)\nonumber\\
&&\otimes(\delta_a|H\rangle_a-\eta_a|V\rangle_a)(\delta_b|H\rangle_b+\eta_b|V\rangle_b) \nonumber\\
&&\otimes\cdots\otimes(\delta_z|H\rangle_z+\eta_z|V\rangle_z).
\;\;\;\;\;\;\;\;\;\;\;\;\;\;\;\;\;\;\;\;\;\;\;\;\;\;\;
\label{N-initial3}
\end{eqnarray}

In the decoding procedure, with the decoder shown in Fig. \ref{fig3} (b), the parties Alice,
Bob, $\cdots$, and Zach let the photons $a$, $b$, $\cdots$, and $z$
pass through their   unbalanced polarization interferometers
followed by a PC. PBS$_{11}$ followed with a time delay $\Delta t$ on
the $|V\rangle$ components is used to separate the $|H\rangle$ and
$|V\rangle$ components of the photon and let them  pass through
QWP$_1$ and QWP$_2$, respectively. After the photons successively
pass through the optical elements described above,  the  $N$-photon
state evolves into
\begin{eqnarray}
|\Phi^-_N\rangle_{t_2}\!\!&=&\!\!
\frac{1}{\sqrt{2}}(|R^{\uparrow}\rangle_a|R^{\uparrow}\rangle_b\cdots|R^{\uparrow}\rangle_z
\!-\!|L^{\downarrow}\rangle_a|L^{\downarrow}\rangle_b\cdots|L^{\downarrow}\rangle_z)\nonumber\\
&&\otimes[(\delta_a|s'\rangle_a+\eta_a|l'\rangle_a)
(\delta_b|s'\rangle_b+\eta_b|l'\rangle_b)\nonumber\\
&&\otimes\cdots\otimes(\delta_z|s'\rangle_z+\eta_z|l'\rangle_z)].
\end{eqnarray}
Here $|s'\rangle\equiv|ss\,l\rangle$, $|s\,ls\rangle$, or
$|lss\rangle$ and $|l'\rangle\equiv|s\,l\,l\rangle$,
$|ls\,l\rangle$, or $|l\,ls\rangle$.

If all the spins are initialized to be a superposition state of the
form
$|\Phi\rangle_{e_i}=\frac{1}{\sqrt2}(|\uparrow\rangle+|\downarrow\rangle)$,
 and the photons are in the state
$|\Phi^-_N\rangle_t=\frac{1}{\sqrt{2}}(|R^{\uparrow}\rangle_a|R^{\uparrow}\rangle_b\cdots|R^{\uparrow}\rangle_z
-|L^{\downarrow}\rangle_a|L^{\downarrow}\rangle_b\cdots|L^{\downarrow}\rangle_z)|s'\rangle_a|s'\rangle_b\cdots|s'\rangle_z$,
the state of the hybrid system composed of the $N$ photons and the
$N$ electron spins after their interaction assisted by QD-cavity
systems can be described as
\begin{eqnarray}
|\Phi_{_H}\rangle\!\!&=&\!\!\! \frac{1}{\sqrt{2^{N-1}}}
\sum_{\alpha_{_z}=0}^{1}\!\cdots\!\sum_{\alpha_b=0}^1\!\sum_{\alpha_a=0}^1
-(-1)^\lambda\bigg\{\big[\prod_{i=a}^z(\sigma_{x_i})^{\alpha_i} \nonumber\\
&&\!\!\!-\prod_{i=a}^z(\sigma_{x_i})^{\bar{\alpha_i}}\big]|R^{\uparrow}R^{\uparrow}\cdots{}R^{\uparrow}\rangle_{ab\cdots{}z}
\!\otimes\big[\prod_{i=a}^z(\sigma^i_{x})^{\alpha_i}\nonumber\\
&&\!\!\!-(-1)^N\!\prod_{i=a}^z(\sigma^{i}_{x})^{\bar{\alpha_i}}\big]|\uparrow\uparrow\cdots{}
\uparrow\rangle_{e_ae_b\cdots{}e_z}\!\bigg\},
\end{eqnarray}
where the single-qubit operators
$\sigma_{x_i}=|R^{\uparrow}\rangle_i\langle{}L^{\downarrow}|+|L^{\downarrow}\rangle_i\langle{}R^{\uparrow}|$
and
$\sigma^i_{x}=|\uparrow\rangle_i\langle\downarrow|+|\downarrow\rangle_i\langle\uparrow|$
are used to complete the bit-flip operations on the $i$-th photon
and the $i$-th electron spin, respectively, and
$\bar{\alpha_i}=1-\alpha_i$, while the parameter $\lambda=\sum\alpha_i$. Alice,
Bob, $\cdots$, and Zach measure the photons $a$, $b$, $\cdots$, and
$z$, respectively,  in the $\{|R\rangle$, $|L\rangle\}$ basis, and
the electron-spin subsystem will be projected into a maximally
entangled N-spin GHZ state. To be detail, if the collective outcome
of the measurement is
$|R^{\uparrow}R^{\uparrow}\dots{}R^{\uparrow}\rangle_{ab\cdots{}z}$,
the electron-spin subsystem will be collapsed into the state
$|\Phi^+_{_N}\rangle_{e}=\frac{1}{\sqrt2}
(|\uparrow\uparrow\dots{}\uparrow\rangle-(-1)^N|\downarrow\downarrow\dots{}\downarrow\rangle){e_ae_b\cdots{}e_z}$.
If the number $N$ of the parties is odd, no additional operation is
required to obtain the target entangled GHZ state
$|\Phi^+_{_N}\rangle_{e_0}=\frac{1}{\sqrt2}
(|\uparrow\uparrow\dots{}\uparrow\rangle+|\downarrow\downarrow\dots{}\downarrow\rangle){e_ae_b\cdots{}e_z}$;
otherwise, Alice performs a phase-flip operation
$\sigma^a_{z}=|\uparrow\rangle_a\langle\uparrow|-|\downarrow\rangle_a\langle\downarrow|$
on the spin $e_a$ to get the target entangled GHZ state
$|\Phi^+_{_N}\rangle_{e_0}$.

During the entanglement distribution process, the quantum noise  on
the polarization mode of the photons in our protocol is general. If
the giant circular birefringence induced by the single electron spin
is reliable, one can complete the entanglement distribution process
and get the $N$ remotely separated QD-confined electron spins
entangled in the GHZ state $|\Phi^-_{_N}\rangle_{e_0}$ in a heralded
way conditioned on the detecting of one photon in each node. The
photon loss during the entanglement distribution process cannot lead to the heralded results and  does not
affect the fidelity of the entangled N-QD-electron-spin states.

\end{document}